%Paper: hep-ph/9207265
%From: collins@phys.psu.edu (John Collins)
%Date: Mon, 27 Jul 92 13:20:51 EDT
%Date (revised): Mon, 27 Jul 92 13:51:06 EDT
%Date (revised): Mon, 27 Jul 92 14:12:22 EDT

% e-mail queries on this paper to collins@phys.psu.edu
%
% This is polfact.tex.
%It is in plain TeX and contains all necessary macro files, except for
%figures.  The full paper, with figures, is obtainable by anonymous ftp
%(Username anonymous ftp) from ftp.phys.psu.edu (in directory
%pub/preprint).
%
%
%
%
%
%========================================
%====== polfact.tex ======
%
%  References corrected 7 July 92.
%
%******************************************************************
%
%\input jccmacro
%========================================
%====== jccmacro.tex ======
%
\catcode`\^^?=9%  Cat code for delete => ignore to handle DOS files under UNIX.
%TeX macros by John Collins.            Revised:  14 May 92
%
%   Symbols moved to jccsym.tex
%
%==========================================================
%   Conditionals and utilities:
%
\def\ifundef#1{\expandafter\ifx\csname #1\endcsname\relax}
\def \IFX #1=#2\THEN #3\ELSE #4\ENDIF
   {\edef \TMPIFA {#1}\relax \edef \TMPIFB {#2}\relax \ifx \TMPIFA
\TMPIFB #3\else #4\fi }%
\def \IFUNDEF #1\THEN #2\ELSE #3\ENDIF
   {\ifundef {#1}#2\else #3\fi
}%
%
% absorbspace: remove whitespace.
% Warning: There must be no whitespace in macro definition!!
%
\long\def \captureCR #1{\def \abspCR {#1}}
\captureCR
\long\def \absorbspace#1{\def \abspARG{#1}\def \abspW{ }%
\def \abspRES{#1}% Default result if #1 is not white.
\ifx \abspARG \abspW \def \abspRES {\absorbspace}\fi %
\ifx \abspARG \par \def \abspRES {\absorbspace}\fi %
\ifx \abspARG \abspCR \def \abspRES {\absorbspace}\fi %
\abspRES}
%
%
%====================================
%   Symbols:
%
\def\frac#1#2{{#1\over #2}}
%\input jccsym
%========================================
%====== jccsym.tex ======
%
%      Math/physics symbols.
%      Revised by J. Collins 8 May 92
%
% This file is aimed to be same for both plain TeX and LaTeX.
% For plain TeX, \frac should have been defined earlier.
%
%\def\ifundef#1{\expandafter\ifx\csname #1\endcsname\relax}
%\newif \ifLaTeX
%\ifundef{newcommand}\LaTeXfalse \else \LaTeXtrue \fi
%=================================================================
%
%  OBSOLETE:
%
% \def\pr#1{#1^\prime} % OVERRIDDEN BY PHYS REV
%
%=================================================================
%
%
% This defines fractions with appropriate sizes:
%??  Old version: \def\fr#1#2{{\textstyle{#1 \over #2}}}
\def \fr#1#2{\ifinner \innerfr{#1}{#2}  \else \outerfr{#1}{#2}\fi}
\def \innerfr#1#2{\ifmmode \ndmathfr{#1}{#2}
           \else \textfr{#1}{#2}\fi}
\def \outerfr#1#2{\ifmmode \dmathfr{#1}{#2}
           \else \textfr{#1}{#2}\fi}
\def \ndmathfr #1#2{ \raise 0.4 ex\hbox{${\scriptstyle {#1 \over #2}} $} }
\def \dmathfr  #1#2{ {#1 \over #2} }
\def \textfr   #1#2{\raise 0.4 ex\hbox{${\scriptstyle {#1 \over #2}}$}}
%
%
% Sub/superscript preceeding current position:

%
% Isotope symbol:

%
%
%=================================================================
%
%
%

\def\fun#1#2{\lower3.6pt\vbox{\baselineskip0pt\lineskip.9pt
  \ialign{$\mathsurround=0pt#1\hfil##\hfil$\crcr#2\crcr\sim\crcr}}}
%approx less than and greater than
\def\centeron#1#2{{\setbox0=\hbox{#1}\setbox1=\hbox{#2}\ifdim
\wd1>\wd0\kern.5\wd1\kern-.5\wd0\fi
\copy0\kern-.5\wd0\kern-.5\wd1\copy1\ifdim\wd0>\wd1
\kern.5\wd0\kern-.5\wd1\fi}}
\def\centerover#1#2{\centeron{#1}{\setbox0=\hbox{#1}\setbox
1=\hbox{#2}\raise\ht0\hbox{\raise\dp1\hbox{\copy1}}}}
\def\centerunder#1#2{\centeron{#1}{\setbox0=\hbox{#1}\setbox
1=\hbox{#2}\lower\dp0\hbox{\lower\ht1\hbox{\copy1}}}}
%
%===================================================
%
%     MAIN DEFINITIONS:
%
%----------------These define et al., i.e., e.g., cf., etc:

%
%----------------Units:  ====
%  Some have Greek in them, so swithc to mathmode if needed:
\def \UNIT #1{{\ifmmode \UNITBODY {#1} \else $ \UNITBODY {#1} $\fi }}
\def \UNITBODY #1{\,{\rm #1}}

%
%------------Common physics symbols
\def \al {\alpha_{\rm s}}

\def \tr{\mathop{\rm tr}}

% transverse vector

%\def \MSbar {\vbox{\hrule\kern 1pt\hbox{\rm MS}}}
\def \MSbarbasic {\overline{{\rm MS}}}
\def \MSbar {\ifmmode \MSbarbasic \else $\MSbarbasic$\fi }

      % fancy L for the Lie derivative
%       The command \st (for stroke) puts a slash through the succeeding
%       character in math mode

\def \bra#1|{\mathopen{\langle#1\,|}}  %Dirac bra arg. is placed in bra symbol
\def \braket#1|#2>{\langle#1\,|\,#2\rangle} %Dirac braket arg's in bra & ket
\def \ket#1>{\mathclose{|\,#1\rangle}}  %Dirac ket arg. is placed in ket symbol

\def \ltap{\;\centeron{\raise.35ex\hbox{$<$}}{\lower.65ex\hbox{$\sim$}}\;}
\def \gtap{\;\centeron{\raise.35ex\hbox{$>$}}{\lower.65ex\hbox{$\sim$}}\;}

\def \d{{\rm d}}

\def \e{{\rm e}}
\def \epem{{\ifmmode \epembody \else $ \epembody $\fi }}
%  Allow correct size in sub and super.
\def \epembody{ \e^+\e^- }
\def \half{\fr12}

\def \P{{\rm P}}
%
%------------------Journals:
\def \rf    #1#2#3{#1, #2 (19#3)}
\def \jour  #1{#1\ \relax}

\def \np    {\jour {Nucl.\ Phys.}}

\def \pl    {\jour {Phys.\ Lett.}}

\def \pr    {\jour {Phys.\ Rev.}}

\def \prl   {\jour {Phys.\ Rev.\ Lett.}}

\def \zphc  {\jour {Z.\ Phys.\ C}}

\relax
%
%====== END OF jccsym.tex ======
%=========================================
%\input fullxref
%========================================
%====== fullxref.tex ======
%
%
%   xref macros
%
%   ***************************  JCC  20 Sep 89 *************
%
%==========================================
%
%         DEFINE REFERENCE FORMAT:
%

\def \ref[#1]{[\dorefs#1,/]}
\def \Ref[#1]{ref.~[#1]}
\def \Refr[#1]{\dorefs#1,/}
\def \defref#1#2{\bibitem{#1}{#2}\par}
%
%==========================================
%
%         EQUATIONS:
%
\def \eqref#1{#1}
\def \er(#1){(\eqref{#1})}
\def \eq(#1){eq.~\er(#1)}
\def \Eq(#1){Equation \er(#1)}
\def \eqalno(#1){(#1)}
%
%============================================
%
%  FIGURES defined in JCCMACRO.TEX or JCCART.STY
%
\relax

%
%====== END OF fullxref.tex ======
%=========================================
%
%-----------------------------------------------------------------
%                      DATE:
\edef\date {\ifcase\month\or
  January\or February\or March\or April\or May\or June\or
  July\or August\or September\or October\or November\or December\fi
  \space\number\day, \number\year}
%
%-----------------------------------------------------------------
%               MISCELLANEOUS COMMANDS:
%
%                   New page:

%
%               Discretionary break, no hyphen:

%
%               Set draft mode etc:

\newif \ifdraft
\draftfalse
%
%              Underline using no mathmode.
\def\undertext#1{\vtop{\hbox{#1}\kern 1pt \hrule}}
%
%              Centering:
\def \centered {\leftskip=0pt plus 1fill \rightskip=\leftskip }
%
%              Eq
\def\eqn\par#1\par{$$#1$$}
%
%-----------------------------------------------------------------
%             FIGURES:

\def \beginfig #1{\begingroup \singlespacing
            \narrower
            \midinsert \vskip 0.5 in \vskip #1 true in
            \noindent \lineskip=1pt}
\def \endfig {\smallskip \endinsert \endgroup}

%
%        Format of figure caption.  Also useful for building up
%        complicated figure.
\def \deffigcap#1#2{Fig.~#1.  #2}
\def \deffig#1#2#3{\beginfig {#2}
%        Figure with caption.
%        #1 = figure number,
%        #2 = height of figure in true inches,
%        #3 = caption.
   \noindent \deffigcap{#1}{#3}
   \endfig}
\def \fig#1{fig.~#1}
\def \Fig#1{Fig.~#1}
%
%
%======================================================================
%
%        POSTSCRIPT FIGURES: EXPERIMENTAL
%

%

%
\def\PrintChart#1#2#3#4#5#6{
%       #1 = postscript file.
%       #2 = height of TeX box, in inches, before scaling
%       #3 = width of TeX box, in inches, before scaling
%       #4 = Translation to left, to apply to figure before scaling.
%       #5 = Translation down, to apply to figure before scaling.
%       #6 = scaling factor
   \vskip 10 pt \nobreak
   \hbox to \hsize{%
   \hss
   \dimen0=#3in%
   \hbox to #6\dimen0{%
      \dimen0=#2in%
      \vbox to #6\dimen0{
         \vss
          \special{ps: plotfile #1 asis}
          \special{ps::[asis,end]
             ChartCheckPoint restore
             0 SPE
          }
       }%
       \hss
    }
    \hss
    }
    \vskip 10 pt
}
%
   % The sizing doesn't put the figure in the correct place
%
%
%-----------------------------------------------------------------
%                    SPACING MACROS
%    NEED TO FIX THESE UP TO WORK WITH twelvepoint and tenpoint
\def \singlespacing{\baselineskip=11pt}
\def \doublespacing{\baselineskip=22pt}
%
%-----------------------------------------------------------------
%                   SIZE: 10 and 12 pt
%
%
% twelve-point macro (from Barry) for new amr fonts
% tw.tex  for TEX 1.0
%
% NOTE: \twelvepoint \tenpoint
%    these commands will change \baselineskip retroactively
%    for the whole of the 'current' paragraph and after.
%    So unless you want 'special effects',
%    you should put \twelvepoint and \tenpoint after a blank line to force
%    termination of the previous paragraph if you are changing between 10 & 12.
%
% the following set of lines copied from the TEX82 1.1 PLAIN.TEX
% to make this work with TEX82 1.0
%
%  JCC: I have changed these to use 12pt fonts or scaled fonts.  That
%       works better.
%
\font\twelverm=cmr12 % roman text
\font\twelvei=cmmi12 % math italic
\font\twelvesy=cmsy10 scaled \magstep1 % math symbols
\font\twelvebf=cmb10 scaled \magstep 1 %use boldface
\font\twelvesl=cmsl12 % slanted roman
\font\twelveit=cmti12 % text italic
%===\font\twelverm=cmr10 at 12 pt % roman text
%===\font\twelvei=cmmi10 at 12 pt  % math italic
%===\font\twelvesy=cmsy10 at 12 pt  % math symbols
%===\font\twelvebf=cmb10 at 12 pt %use boldface
%===\font\twelvesl=cmsl10 at 12 pt % slanted roman
%===\font\twelveit=cmti10 at 12 pt % text italic
\font\ninerm=cmr9
\font\ninei=cmmi9
\font\ninesy=cmsy9
 %boldface extended for smaller fonts
%
\font\twelvett=cmtt12 % typewriter
\font\twelvebf=cmb10 scaled \magstep1
%===\font\twelvett=cmtt10 at 12 pt  % typewriter
%===\font\twelvebf=cmb10 at 12 pt
\def\twelvepoint{\def\rm{\fam0\twelverm}
\textfont0=\twelverm \scriptfont0=\ninerm \scriptscriptfont0=\sevenrm
\textfont1=\twelvei \scriptfont1=\ninei \scriptscriptfont1=\seveni
\textfont2=\twelvesy \scriptfont2=\ninesy \scriptscriptfont2=\sevensy
\textfont3=\tenex \scriptfont3=\tenex \scriptscriptfont3=\tenex
\def\it{\fam\itfam\twelveit}%
\textfont\itfam=\twelveit
\def\sl{\fam\slfam\twelvesl}%
\textfont\slfam=\twelvesl
\def\bf{\fam\bffam\twelvebf}%
\textfont\bffam=\twelvebf
\def\tt{\fam\ttfam\twelvett} % \tt is family 7
\textfont\ttfam=\twelvett
\baselineskip=14pt%
\abovedisplayskip=14pt plus 3pt minus 10pt%
\belowdisplayskip=14pt plus 3pt minus 10pt%
\abovedisplayshortskip=0pt plus 3pt%
\belowdisplayshortskip=8pt plus 3pt minus 5pt%
\parskip=3pt plus 1.5pt
\setbox\strutbox=\hbox{\vrule height10pt depth4pt width0pt}%
\rm}
\def\tenpoint{\def\rm{\fam0\tenrm}%
\textfont0=\tenrm \scriptfont0=\sevenrm \scriptscriptfont0=\fiverm
\textfont1=\teni \scriptfont1=\seveni \scriptscriptfont1=\fivei
\textfont2=\tensy \scriptfont2=\sevensy \scriptscriptfont2=\fivesy
\textfont3=\tenex \scriptfont3=\tenex \scriptscriptfont3=\tenex
\def\it{\fam\itfam\tenit}%
\textfont\itfam=\tenit
\def\sl{\fam\slfam\tensl}%
\textfont\slfam=\tensl
\def\bf{\fam\bffam\tenbf}%
\textfont\bffam=\tenbf
\def\tt{\fam\ttfam\tentt}%
\textfont\ttfam=\tentt
\baselineskip=12pt%
\abovedisplayskip=12pt plus 3pt minus 9pt%
\belowdisplayskip=12pt plus 3pt minus 9pt%
\abovedisplayshortskip=0pt plus 3pt%
\belowdisplayshortskip=7pt plus 3pt minus 4pt%
\parskip=0.0pt plus 1.0pt
\setbox\strutbox=\hbox{\vrule height8.5pt depth3.5pt width0pt}%
\rm}
%
%-----------------------------------------------------------------
%                    SECTION HEADING MACROS
%
%
\def \mainhead #1{\goodbreak \bigskip \bigskip
        \vbox{\noindent\raggedright\uppercase{#1}}
         \nobreak \medskip \nobreak
         \par \noindent \absorbspace }
\def \sechead #1{\mainhead {\number\secnum.  #1}}
\def \subsechead #1{\medskip \goodbreak
            \leftline{\it \number\secnum.\number\subsecnum~~#1} \nobreak
\smallskip \noindent}
\def\subsubsechead#1{\goodbreak \smallskip \noindent {\uppercase{#1~~}}}
\newcount\secnum \secnum=0
\newcount\subsecnum \subsecnum=0
\newcount\subsubsecnum \subsubsecnum=0
% The following must have no spaces after the heading, else noindent in
%    heading goes wrong:
\def\section#1{\global\advance\secnum by 1
            \subsecnum=0
            \immediate\write0{=== \the\secnum\ #1}
            \sechead{#1}}
\def\subsection #1{\global\advance\subsecnum by 1 \subsubsecnum=0%
            \subsechead{#1}}
\def\subsubsection #1{\global\advance\subsubsecnum by 1 \subsubsechead{#1}}
\def \sec #1\par{\section {#1}}
\def \subsec #1\par{\subsection {#1}}
\def \subsubsec #1\par{\subsubsection {#1}}
%
%-----------------------------------------------------------------
%                       REFERENCE MACROS
\newif\ifuserefnums
\userefnumstrue

\def\ref[#1]{\ [\dorefs#1,/]}
\def\onebibitem #1#2{\item{#1}#2\par}
\newcount\numrefs  \numrefs=0
\newcount\nextref
\def\dorefs#1,#2/{\doref{#1}%
{\def\a{#2}\def\b{}\ifx\a\b\relax\else, \dorefs#2/\fi}}
\def\doref#1{\expandafter\ifx\csname refnum#1\endcsname\relax%
\global\advance \numrefs by 1%
\expandafter\xdef\csname refnum#1\endcsname{\number\numrefs}%
\expandafter\xdef\csname ref\number\numrefs\endcsname{#1}%
\expandafter\xdef\csname reftext#1\endcsname{Ref.\ #1 not yet defined.}%
\fi%
\ifuserefnums\csname refnum#1\endcsname%
\else #1\fi}
%
%==== \boxref is used when citing literature in an item -- for example
%     in a \defref -- for direct use of \ref causes wierd TeX errors
%     Usage like \boxref \refname \refbox {\ref[PMS]}  outside items.
%           invoke by \refname.  \refbox is used to hold the
%           reference.
% This is the plain TeX version, with a fudge to allow \newbox in the
% macro definition: Plain TeX does not allow \newbox to appear
% explicitly in a macro definition.
\def\boxref#1#2#3{\NEWBOX #2
                  \setbox #2=\hbox {#3}
                  \def #1{\unhcopy #2}}
\let \NEWBOX=\newbox
\def\bibitem#1#2\par{\unskip\expandafter\ifx\csname refnum#1\endcsname\relax%
        \expandafter\xdef\csname reftext#1\endcsname{{\bf Uncited: }#2}%
        \immediate\write0{*** \onebibitem{#1}{\csname reftext#1\endcsname}   }
      \else%
        \expandafter\xdef\csname reftext#1\endcsname{#2}%
      \fi
      \ifuserefnums\relax\else\onebibitem{#1}{\csname reftext#1\endcsname}\fi}
\def \defref#1#2{\bibitem{#1}{#2}\par}
\def\outrefs{\ifuserefnums{\nextref=0%
   \loop%
     \ifnum\nextref<\numrefs%
     \advance\nextref by 1%
     \onebibitem{\number\nextref}{\csname reftext\csname
                    ref\number\nextref\endcsname\endcsname}%
   \repeat}\fi}
\def \refstring[#1]{\ifuserefnums\csname refnum#1\endcsname \else #1\fi}
%
%
%====== END OF jccmacro.tex ======
%=========================================
%===\labelrefs
\twelvepoint
%
% Abbreviation for path ordered exponential:
\def \P {\ {\cal P}\ }
\vskip 1.0in
\rightline {PSU/TH/100}
\rightline {May 19, 1992}
\vskip 1.0in

{\centered
\obeylines

   {\bf HARD SCATTERING IN QCD WITH POLARIZED BEAMS }

      \bigskip

   John C. Collins\footnote*{E-mail: collins@phys.psu.edu or
                            BITNET: collins@psuleps.}
       \smallskip
     Physics Department
     Pennsylvania State University
     State College, PA 16802, U.S.A.
}

\doublespacing

\bigskip
\bigskip

{\narrower
\singlespacing

            \centerline{ABSTRACT}

    I show that factorization for hard processes in QCD is also
    valid when the detected particles are polarized, and that the
    proof of the theorem determines the operator form for the
    parton densities.  Particular attention is given to the case
    of transversely polarized incoming hadrons.

}

%=====================================================================
\section {Introduction}%
Perturbative QCD is by now a staple ingredient of most
phenomenology of high energy collisions.  Mostly the applications
have been to the case that the incoming particles are
unpolarized.  However, in recent years much more attention has
been devoted to the polarized case. Unfortunately, there has been
confusion as to the exact status of the factorization theorem in
the polarized case.  This is the theorem that underlies almost
all perturbative calculations.  The confusion has particularly
extended to the question of the correct definition of the parton
distribution functions\ref[controv,F,IKL].

In actuality, when one examines the proofs of the factorization
theorem\ref[CSS,CSS1], one finds that almost no reference is
made to the polarization of the measured particles. So the
purpose of this paper is to explain that the factorization
theorem does indeed apply, and that it determines, essentially
uniquely, the definition of the parton distributions. (The
ambiguity in the definition is directly tied to the well-known
freedom to choose the renormalization and factorization scales,
and is entirely independent of the problems of polarization.)
There has been particular confusion about the case about the case
that the incoming beams have transverse polarization.  For
example of the supposed problems, see the discussion in the book
by Ioffe et al,\ref[IKL].  Better and more recent discussions of
some of the issues can be found in \ref[RS,AM,PSU,JJ,QS,BB].

The processes with which we are concerned are those with a hard
scattering: deeply inelastic lepton scattering, the Drell-Yan
process, jet production in hadron collisions, etc.  For the sake
of definiteness, I will treat in this paper two specific cases:
the structure functions for deeply inelastic lepton scattering,
and the Drell-Yan process.

All the pieces of the discussion can be found in the literature.
What has been lacking is a unified presentation.  In this paper,
I shall be concerned solely with the `twist-2' behavior of
hard scattering cross sections.  This means that at each order of
perturbation theory I consider the terms that scale with energy
like their dimension (modified by logarithms).  `Higher twist'
terms---power suppressed terms---will be ignored; they are very
interesting, but their study is much harder than that of the
twist-2 terms.

The terminology of twist has its origins in the operator product
expansion for deeply inelastic lepton scattering, and is somewhat
inappropriate here.  But the usage has stuck.

%=====================================================================
\section {Statement of Factorization Theorem}%
In this section, I will formulate the factorization theorems for
deeply inelastic scattering and for the Drell-Yan cross section.
These are typical of the most general case of factorization.

%------
\subsection {Structure Functions for Deeply Inelastic Scattering}%
The structure functions for deeply inelastic scattering are
defined in terms of the structure tensor $W_{\mu \nu }(p,q)$ by:
$$
\eqalign{
  W_{\mu \nu } = & \left(-g_{\mu \nu }+q_{\mu }q_{\nu }/q^{2}\right)
W_{1}(x,Q^{2})
        + \left(p_{\mu }-q_{\mu }p\cdot q/q^{2}\right) \left(p_{\nu }-q_{\nu
}p\cdot q/q^{2}\right)
          \frac {W_{2}(x,Q^{2})}{M^{2}}
\cr
        &+ \frac {i}{M} \epsilon _{\mu \nu \rho \sigma }q^{\rho } \left[
s^{\sigma }\left( G_{1}+\frac {p\cdot q}{M^{2}} G_{2} \right) -s\cdot q
p^{\sigma }\frac {1}{M^{2}}G_{2} \right].
}
\eqno(1)
$$
Here, $M$, $p^{\mu }$ and $s^{\mu }$ are the mass, momentum and spin vector
of the target, and $q^{\mu }$ is the momentum of the exchanged virtual
photon. (The case of a more general exchanged boson, like a W or
a Z gives more structure functions, a situation that is an
inessential complication for the present purpose.)  As usual, we
define $Q^{2}=-q^{2}$, $\nu =p\cdot q$, and $x=Q^{2}/2p\cdot q$.  We will be
interested
in the Bjorken limit: $Q\to \infty $ with $x$ fixed.  Our normalization of
the spin vector is such that a pure state has $s^{2}=-1$.

As for the spin-dependent structure functions, we will consider
in this paper only the case that the target has spin \half.  Then
target's spin state (no matter whether mixed or pure) is
completely determined by its spin vector $s^{\mu }$,  and there are
exactly two polarized structure functions, $G_{1}$ and $G_{2}$.  (For
the case of a target of general spin, see \ref[HighSpin].)  The
normalization of $s^{\mu }$ is that it satisfies $0\geq s^{2}\geq -1$ and
$s\cdot p=0$,
and that a pure spin state has $s^{2}=-1$.

Later, when we do power counting to determine the sizes of the
leading contributions to the structure functions, it will be
convenient to work in the center-of-mass frame.  Then in the
Bjorken limit, we find that components of $p^{\mu }$ and of $q^{\mu }$ are of
order $Q$.  But also the components of $s^{\mu }$ become large. So,
following Ralston and Soper\ref[RS], we decompose
$s^{\mu }$ in terms of a helicity $\lambda $ and a
transversity $s_{\perp }$:
$$
   s^{\mu } = \lambda  \left( \frac {p^{\mu }}{M} - \frac {q^{\mu }M}{p\cdot q}
\right) \frac {1}{\sqrt {1+2xM^{2}/p\cdot q}} \, + {s_{\perp }}^{\mu },
\eqno(2)
$$
where the helicity $\lambda $ is defined by
$$
   \lambda  = \frac {q\cdot sM}{q\cdot p}\,  \frac {1}{\sqrt {1+2xM^{2}/p\cdot
q}},
\eqno(3)
$$
and $s_{\perp }^{\mu }$ is orthogonal to both $p^{\mu }$ and $q^{\mu }$. Then
$-s^{\mu }s_{\mu }=\lambda ^{2}+|s_{\perp }|^{2}$, while both $|\lambda |$ and
$|s_{\perp }|$ are less than
unity. For a pure state, $\lambda ^{2}+|s_{\perp }|^{2}=1$.

It is common to call $s_{\perp }$ the transverse spin.  However, moving
particles are not in an eigenstate of transverse spin, as Jaffe
and Ji\ref[JJ] explained, but may be in an eigenstate of the
transverse components of the Pauli-Lubanski vector; these
transverse components are called transversity.  Note that our
spin vector $s^{\mu }$ is proportional to the Pauli-Lubanski vector.)
Polarized particles in a high energy accelerator are typically
transversely polarized and carry a definite value of the spin
vector.  Thus they are in a state of definite transversity. Note
that the concept of `transverse' in this context is {\it not}
Lorentz invariant, when referred to a single particle.  It is
only defined when one brings another vector into the situation to
represent (say) the center-of-mass of the scattering.

At various stages, we will need to take the limit of zero mass,
and the decomposition \er(2) exhibits potential singularities at
$M=0$. In this limit, we may approximate \eq(2) by
$$
   s^{\mu } = \lambda  \frac {p^{\mu }}{M} + {s_{\perp }}^{\mu } + {\rm power\
law\ correction},
\eqno(4)
$$

The factorization properties are most easily expressed in terms
of scaling structure functions which are defined by rewriting
\eq(1) as
$$
\eqalign{
  W_{\mu \nu } = & (-g_{\mu \nu }+q_{\mu }q_{\nu }/q^{2}) F_{1}(x,Q^{2})
        +  \frac {(p_{\mu }-q_{\mu }p\cdot q/q^{2}) (p_{\nu }-q_{\nu }p\cdot
q/q^{2})}{p\cdot q} F_{2}(x,Q^{2})
\cr
        &+ \frac {iM}{p\cdot q} \epsilon _{\mu \nu \rho \sigma }q^{\rho
}s^{\sigma } g_{1}
         + \frac {iM}{(p\cdot q)^{2}} \epsilon _{\mu \nu \rho \sigma }q^{\rho
}(p\cdot qs^{\sigma } -s\cdot q p^{\sigma }) g_{2}.
%        &+ \frac {i \lambda  \epsilon _{\mu \nu \rho \sigma }q^{\rho
% }p^{\sigma }}{p\cdot q}  G_{L} + \frac {i\epsilon _{\mu \nu \rho \sigma
% }q^{\rho }s^{\sigma }_{\perp }}{p\cdot q}  M G_{T}.
}
\eqno(5)
$$
Here, $F_{1}\equiv W_{1}$, $F_{2}\equiv p\cdot qW_{2}/M^{2}$, $g_{1}\equiv
p\cdot qG_{1}/M^{2}$ and $g_{2}\equiv p\cdot q^{2}G_{2}/M^{4}$.
%============================================================
% ===?? while the spin-dependent structure
%functions are defined by:
%$$
%\eqalign{
%   G_{1} &= \frac {M^{2}}{p\cdot q \sqrt {1+2xM^{2}/p\cdot q}} G_{L} + \frac
% {2x M^{4}}{p\cdot q^{2} (1+2xM^{2}/p\cdot q)} G_{T},
%\cr
%   G_{2} &=  \frac {M^{4}}{p\cdot q^{2} (1+2xM^{2}/p\cdot q)} G_{T} - \frac
% {M^{4}}{p\cdot q^{2} \sqrt {1+2xM^{2}/p\cdot q}} G_{L} ,
%}
%\eqno(5)
%$$
%============================================================

In the Bjorken limit, it is convenient to use the decomposition
\eq(2) of the spin, so that the spin-dependent part of $W_{\mu \nu }$ is:
$$
\eqalign{
  W^{{\rm pol}}_{\mu \nu } =
        & \lambda  \frac {i\epsilon _{\mu \nu \rho \sigma }q^{\rho }p^{\sigma
}}{p\cdot q}  \frac {1}{\sqrt {1+2xM^{2}/p\cdot q}}
            \left(g_{1} - \frac {2xM^{2}}{p\cdot q}g_{2} \right)
\cr
        &+ \frac {i\epsilon _{\mu \nu \rho \sigma }q^{\rho }s_{\perp }^{\sigma
}M}{p\cdot q} g_{2}
\cr
   = & \lambda  \frac {i\epsilon _{\mu \nu \rho \sigma }q^{\rho }p^{\sigma
}}{p\cdot q}  g_{1} + \lambda  \frac {i\epsilon _{\mu \nu \rho \sigma }q^{\rho
}s_{\perp }^{\sigma }M}{p\cdot q} g_{2}
     + \hbox{non-leading powers}.
}
\eqno(6)
$$

We will see that in the Bjorken limit, $Q\to \infty $ with $x$ fixed, each
of $F_{1}$, $F_{2}$, $g_{1}$, and $g_{2}$ scales like $Q^{0}$ times logarithms.
Each of these scaling structure functions is dimensionless and,
except for $g_{2}$, its definition in terms of the tensor does not
involve the mass of the target.  The exception for $g_{2}$ is a
choice that directly reflects the fact that its leading
contribution is associated with operators of twist-3 rather than
twist-2: redefining it to remove the factor of $M$ would
reduce its power law by one power of $Q$.

% OLD VERSION WITH GT AND GL=========================
%To leading power in $Q$, the relation between the two sets of
%polarized structure functions is given by:
%$$
%\eqalign{
%   G_{1} &= \frac {M^{2}}{p\cdot q} G_{L} + \hbox {power law correction},
%\cr
%   G_{2} &=  \frac {M^{4}}{p\cdot q^{2}} (G_{T} - G_{L}) + \hbox {power law
% correction}.
%}
%\eqno(6)
%$$
%============================================================

%------
\subsection {Factorization for Unpolarized Deeply Inelastic
Scattering}%
The factorization theorem \ref[CSS]
for deeply inelastic lepton scattering
applies in the Bjorken limit, and for the unpolarized structure
functions it gives:
$$
\eqalign{
   F_{1}(x,Q^{2}) &= \sum _{a}
      \int _{x}^{1} \frac {\d \xi }{\xi } \ f_{a/A}(\xi ,\mu )\
       H_{1a} \left( \frac {x}{\xi }, \frac {Q}{\mu }, \al(\mu ) \right)
       + \hbox{remainder},
\cr
   {1 \over x} F_{2}(x,Q^{2}) &= \sum _{a}
      \int _{x}^{1} \frac {\d \xi }{\xi } \ f_{a/A}(\xi ,\mu )\ \frac {\xi }{x}
       H_{2a} \left( \frac {x}{\xi }, \frac {Q}{\mu }, \al(\mu ) \right)
       + \hbox{remainder}.
}
\eqno(7)
$$
This theorem asserts that in the Bjorken limit, the target may be
regarded as a beam of partons, and that the scattering really
takes place on these partons.  The remainder terms are a power of
$Q$ smaller than the leading terms.

The quantities $f_{a/A}(\xi )$ are parton distribution functions (or
parton densities).  Their operator definition, which will be
given below, can be interpreted in light front quantization as
the number density of partons as a function of the light-cone
fraction of the momentum of the parent hadron.

The functions $H_{1a}$ and $H_{2a}$ should be regarded as the
short-distance part of the structure functions for a parton
target of flavor $a$ (gluon, quark or antiquark). In lowest order
in $\al$, each $H_{ia}$ is a delta function at $\xi =x$ times the
charge squared $e_{a}^{2}$ of the parton:
$$
\eqalign{
   H_{1a}(x/\xi ) & = \frac {1}{2} e_{a}^{2} \delta (x/\xi -1) + O(\alpha
_{s}),
\cr
   H_{2a}(x/\xi ) & = e_{a}^{2} \delta (x/\xi -1) + O(\alpha _{s}).
}
\eqno(8)
$$
The first term in these perturbation expansions gives the parton model
approximation to QCD, with $F_{1}=\frac {1}{2}\sum _{a}e_{a}^{2}f_{a/p}(x)$ and
$F_{2}=\sum _{a}e_{a}^{2}xf_{a/p}(x)$.  In higher order, the $H_{ia}$ are the
structure
functions at the parton level, but with subtractions made
according to a standard prescription, to remove the
non-ultraviolet contributions.

The appearance of the Dirac delta function in \eq(8) (and of more
complication generalized functions in higher orders) implies that
the asymptotic behavior given by \eq(7) is to be interpreted in
the sense of distribution theory-- cf.\ \ref[Tk1].

The factors of $1/x$ and of $\xi /x$ in the equation for $F_{2}$ arise
from the dependence on the target momentum of the definition of
the structure function $F_{2}$.

%----
\subsection {Structure Functions v.\ Parton Distributions}%
I make a clear distinction between the concepts of a `structure
function' and a `parton distribution function'.  A structure
function is a term in a decomposition of the deep inelastic
cross section, as in \eq(5); it is experimentally measurable.  A
parton distribution function (or parton density) is a number
density of quarks (or gluons) in a fractional momentum variable.
Mathematically it is a hadron expectation value of a certain
operator; as such, it is a theoretical construct.

However, the parton model gives the structure functions in terms
of certain simple linear combinations of quark and antiquark
densities.  [For our purposes, the parton model is the
(useful) approximation in which one neglects the perturbative
corrections to the hard scattering coefficients in the
factorization formulae.]  It has therefore become common in the
literature to identify the concepts of structure function and
parton distribution.

This identification has particularly disastrous consequences for
the discussion of polarized scattering with transversely
polarized hadrons:  The transverse spin
contribution to deep inelastic scattering at the level of
approximation of the leading term in \eq(7) is exactly zero (to
all orders of perturbation theory), as we will review below.
Attempts to identify the structure function $g_{2}$ with some kind
of transverse spin distribution result in inconsistencies.

%------
\subsection {Factorization for Drell-Yan}%
The factorization theorem for the Drell-Yan process is typical of
factorization theorems for more general hard scattering
processes, and it is formulated as follows.

The process is the inclusive production of a lepton pair of high
invariant mass via an electroweak particle.  The classical case
is with a high-mass virtual photon: $H+H \to  \gamma ^{*} + \hbox{anything}$,
with $\gamma ^{*}\to e^{+}e^{-}$ or $\gamma ^{*}\to \mu ^{+}\mu ^{-}$.  The
cases of $W$ and $Z$ production
can be treated in an essentially identical fashion.

We let $s$ be the square of the
total center-of-mass energy and $q^{\mu }$ be the momentum of the $\gamma
^{*}$.
The kinematic region to which the theorem applies is where $\sqrt s$
and $Q$ get large in a fixed ratio.  ($Q$ is $\sqrt {q^{2}}$.)  The
transverse momentum $q_{\perp }$ of the $\gamma ^{*}$ is either of order $Q$ or
is
integrated over.

In the case that $q_{\perp }$ is integrated over, the factorization
theorem for the unpolarized Drell-Yan cross section reads:
$$
\eqalign{
  \frac {\d\sigma }{\d Q^{2} \d y \d\Omega }   =
   & \quad \sum _{a,b} \int _{x_{A}}^{1} {\d\xi _{A}} \,
      \int _{x_{B}}^{1}{ \d\xi _{B}}\  f_{a/A}(\xi _{A},\mu ) \
        H_{ab}\!\left(\frac {x_{A}}{\xi _{A}}, \frac {x_{B}}{\xi _{B}}, \theta
, \phi , Q; \frac {\mu }{Q}, \al(\mu ) \right)
      \, f_{b/B}(\xi _{B},\mu )
\cr
   & + \hbox{remainder} ,
}
\eqno(9)
$$
where $y$ is the rapidity of the virtual photon and $\d\Omega $ is the
element of solid angle for the lepton pair: the polar angles for
this decay are $\theta $ and $\phi $ relative to some chosen axes. The sums
over $a$ and $b$ are over parton species, and we write
$$
  x_{A} = \e^{y} \sqrt {\frac {Q^{2}}{s}} ,\quad
  x_{B} = \e^{-y} \sqrt {\frac {Q^{2}}{s}} .
\eqno(10)
$$

The function $H_{ab}$ is the ultraviolet-dominated hard scattering cross
section, computable in perturbation theory.  It plays the role of a parton
level cross section and is often written as
$$
H_{ab} =  \frac {\d\hat\sigma }{\d Q^{2} \d y \d\Omega } ,
\eqno(11)
$$
where the hat over the $\sigma $ indicates a hard scattering cross
section at the parton level.
The parton distribution functions, $f$, are the same as in deeply
inelastic scattering.  \Fig{1} illustrates the factorization
theorem.

  \deffig{1}{2.3}{Factorization theorem for Drell-Yan cross
  section.}
%  \defpsfigc{1}{2.3}{Factorization theorem for Drell-Yan cross
%  section.}{polf-1.eps}{1000}{1.5}

%------
\subsection {Twist}%
In both of the above factorization theorems, the dependence of
the hard scattering coefficients ($H_{1a}$ etc) on the large
momentum $Q$ is of the form $Q^{p}$ times logarithms of $Q$, where
$p$ is the dimension of the hard scattering coefficient. This is
true for each separate order of perturbation theory in $\al$. For
the coefficient $H_{ab}$ for Drell-Yan we have $p=-4$, and for $H_{1}$
and $H_{2}$ for deeply inelastic scattering we have $p=0$.  The
quantities multiplying the hard scattering coefficients are
dimensionless parton densities.  Immediate consequences are the
standard scaling laws that $F_{1}$ and $F_{2}$ behave like $Q^{0}$ and
that the Drell-Yan cross section $\d\sigma /\d Q^{2}dy$ behaves like $Q^{-4}$
in
the scaling limit, apart from the usual logarithmic scaling
violations.

Terms of this kind, we will label `twist-2' as a generalization
of the usage in the operator-product expansion for
deep inelastic scattering.  (Twist is the dimension minus the
spin of the operators.)  The remainder terms in the factorization
theorems are therefore called higher twist.

When we consider scattering with transversely polarized hadrons,
there are some processes for which the twist-2 term is exactly
zero.  A notorious example is the single transverse spin
asymmetry of high $p_{\perp }$ particle production in hadron-hadron
scattering.  Another case is the structure function $g_{2}$ for
deeply inelastic scattering. For these processes the leading
twist is twist-3, and the corresponding asymmetries
are proportional to some hadronic mass scale divided by $Q$ at
large $Q$.  The choice of the dimensional factor multiplying $g_{2}$
in \eq(5) to be $M/p{\cdot }q^{2}$ rather than $1/p{\cdot }q^{3/2}$ is an
expression of this fact.

%------
\subsection {Longitudinal Polarization}%
The factorization theorems stated above are for unpolarized
incoming hadrons, and they involve an incoherent sum over parton
types. In the case that the incoming hadrons are polarized, the
theorems need generalization.

For the case of longitudinal polarization, the factorization
statements can be readily formulated simply by simply extending
the sum over parton types $a$ (and $b$) to include a sum over
parton helicities.  The unpolarized parton densities will be a
sum over the helicity densities.  (The asserted theorem is still
in need of the proof which will be summarized later.)

For Drell-Yan, the hard-scattering coefficient in \eq(9) should
be treated as a helicity-dependent cross section at the parton
level.

For deeply inelastic scattering, the formulae for $F_{1}$ and $F_{2}$
will remain unchanged, since these structure functions and the
corresponding parton level structure functions are, by
definition, spin independent. But there will be a factorization
formula for the helicity dependent structure function $g_{1}$,
and this will involve the helicity asymmetry of the parton
densities.  Extra polarized structure functions beyond $g_{1}$ will
be needed for the case of a target of spin greater than \half.
\ref[HighSpin]

%------
\subsection {General Polarization, Including Transverse}%
Now, a characteristic of the quantum mechanical theory of spin is
that interference and coherent phenomena occur even in
circumstances where the physics is otherwise classical.  Such is
the case for the factorization of hard processes when the
detected hadrons have a general polarization.  The problem is one
of interference between scattering of partons of different
quantum numbers.

For the flavor quantum numbers of partons there is no such
interference.  For example, we have no contribution to $F_{1}$ from
an interference between scattering on an up quark and a down
quark:
$$
  \langle u+\gamma ^{*}|\hbox{final state}\rangle  \langle \hbox{final state} |
d+\gamma ^{*}\rangle .
\eqno(12)
$$
The reason is that an examination of the flavor of the final
state of the hard scattering is sufficient to determine which
kind of parton initiated the hard scattering.  Alternatively, one
can examine the term in an operator definition of a parton
distribution that would be appropriate to an interference term:
$$
  \langle p| \bar u \cdots d |p\rangle .        \eqno(13)
$$
Quark number conservation forces this to be zero.  The dots
indicate factors that are irrelevant to the flavor structure.

But for spin, there are no such constraints: The final states
that can be produced from the scattering of left-handed quarks
can be the same as the final states
from right-handed quarks, and therefore the
amplitudes can interfere.  To take account of this, we must equip
the partons entering the hard scattering with a spin density
matrix. Therefore the full specification of a parton distribution
is given by the number density of partons together with the
parton's density matrix.  The number density times the density
matrix has linear dependence
on the spin density matrix of the initial hadron.  Longitudinal
polarization gives the special case that the density matrices of
both the parton and the hadron are diagonal in a helicity basis.
(Note that the use of a density matrix allows the state of a
deeply inelastic scattering
parton to be either pure or mixed:  In an inclusive cross
section, where we sum over unobserved parts of the final state, a
parton that enters the hard scattering can be in a mixed state
even when its parent hadron is in a pure state.)

The most general form for the factorization theorem for Drell-Yan is
$$
\eqalign{
  \frac {\d\sigma }{\d Q^{2} \d y \d\Omega }   =
   &  \sum _{a,b} \int _{x_{A}}^{1} {\d\xi _{A}} \,
      \int _{x_{B}}^{1}{ \d\xi _{B}}\  \rho _{a(\alpha \alpha ')/A} f_{a/A}(\xi
_{A},\mu )
\cr
   & \ H_{a(\alpha \alpha ')b(\beta \beta ')}\!\left(\frac {x_{A}}{\xi _{A}},
\frac {x_{B}}{\xi _{B}}, \theta , \phi , Q; \frac {\mu }{Q}, \al(\mu ) \right)
      \, \rho _{b(\beta \beta ')/B} f_{b/B}(\xi _{B},\mu )
\cr
   & + \hbox{remainder} .
}
\eqno(14)
$$
Here $\rho _{i(\alpha \alpha ')/H}$ is the density matrix of partons of flavor
$i$
in hadron $H$, with $\alpha $ and $\alpha '$ being the helicity indices of
the matrix. The density matrix is of course a function of the
same variables $\xi $ and $\mu $ as the number density $f_{i/h}(\xi ,\mu )$.
The
factorization \er(14) differs from the unpolarized case by the
presence of spin density matrices $\rho $ for the partons and by the
dependence of the hard scattering coefficient on the spin
indices, $\alpha $, $\alpha '$, $\beta $, and $\beta '$.  The density matrix of
a
parton is necessarily a linear function of the spin vector of its
parent hadron.

Calculations of amplitudes in perturbation theory are often made
in a helicity basis\ref[hel].  In that case, it is convenient to
work directly with density matrices in a helicity basis. The
indices $\alpha $, $\alpha '$ etc will take on the values $+$ and $-$ (or $L$
and $R$ for left- and right-handed polarization).

Another method is to work with cut Feynman graphs, for the cross
section.  In that case, one uses a polarization sum for initial
state quarks written in terms of the quark's spin vector:
$$
 \frac {1}{2} p\llap / \, \left( 1 - \lambda \gamma _{5} + \gamma _{5}s\llap
/_{\perp } \right).        \eqno(15)
$$
(Our conventions here are those of Bjorken and Drell\ref[BD].)
The density matrix of the quark is then
$$
\frac {1}{2} \pmatrix{
    1+\lambda     & s^{x}+is^{y} \cr
    s^{x}-is^{y} & 1 - \lambda   \cr
  },
\eqno(16)
$$
where we have chosen the z-axis to be along the 3-momentum of
$p^{\mu }$. For the case of a spin-\half{} hadron, the helicity $\lambda _{a}$
of
a quark is proportional to the helicity $\lambda _{A}$ of it parent hadron,
and similarly for the transversity:
$$
\eqalignno{
   \lambda _{a} &= \Delta _{L,a/A} \lambda _{A},      & \eqalno(17)
\cr
   s_{\perp  a} &= \Delta _{T,a/A} s_{\perp  A}.    & \eqalno(18)
}
$$
Here $\Delta _{L}$ and $\Delta _{T}$ are the longitudinal and transverse spin
asymmetries are a quark in a fully polarized hadron.  They are
functions of the variables $\xi $ and $\mu $, of course, and these
asymmetries give the quark densities defined by Jaffe and
Ji\ref[JJ] by weighting by the unpolarized densities:
$$
\eqalignno{
  f_{La/A} &= \Delta _{La/A} f_{a/A},    & \eqalno(19)
\cr
  f_{Ta/A} &= \Delta _{Ta/A} f_{a/A},     & \eqalno(20)
}
$$
Jaffe and Ji use the notation $g_{1a/A}$ instead of $f_{La/A}$, but
this identification invites confusion with the $g_{1}$
structure function, to which it is directly related in the parton
model.  I have also changed Jaffe and Ji's notation $h$ for the
transversity distribution to $f_{T}$ to correspond with $f_{L}$.

We will give the operator definitions of the parton densities
below, exactly as stated by Jaffe and Ji, and we will justify
that these are the correct definitions to use in the
factorization theorems. These densities satisfy linear evolution
equations (Gribov-Lipatov-Altarelli-Parisi) of a similar
structure to the ones in the unpolarized case.

Exactly corresponding considerations apply to gluons.  They also
have two physical polarization states.  (As we will review below,
it is the states of a massless parton that are relevant for the
factorization theorem.)  However, the gluon has spin one, so that
the offdiagonal terms in its density matrix correspond to linear
rather than transverse polarization.  Moreover, an operator
measuring these offdiagonal elements has helicity two, unlike the
operator for quarks, which has helicity one, as can be seen from
the operator definitions below.  Thus it is a
consequence of angular momentum conservation, as explained by
Artru and Mekhfi\ref[AM], that there are no linearly polarized
gluon partons in a spin-\half{} hadron. Furthermore, the
evolution equation for transversely polarized quarks has no
mixing with gluons, and vice versa.

This result depends on the azimuthal symmetry of the operators
measuring the parton densities.  One way of evading the result is
to use a process sensitive to the intrinsic transverse momentum
of the partons \ref[trfr-sgl].

%---
\subsection{Consequences of Chiral Symmetry}%
In QCD there are chiral symmetries that if unbroken would
actually prohibit some of the interference terms that would
otherwise occur with transverse polarization.  At the level of
the hard scattering, as in \er(12), one is working in
perturbation theory with quark masses neglected.  Thus the chiral
symmetries are exact for the hard scattering coefficients at the
twist 2 level.

An important manifestation of this is conservation of quark
helicity in massless perturbation theory.  As we will see below,
a particular consequence of this is that the twist 2 contribution
to the structure function $g_{2}$ is exactly zero.  That is, the
transverse spin asymmetry in ordinary deeply inelastic scattering
is of the order of a hadron mass divided by $Q$ for large $Q$.

But one can easily conceive of hard scattering with two partons
in the initial state, and then there is no reason for the
interference terms to vanish\ref[RS].

At first glance, the same reasoning might appear to apply to the
parton densities, as defined by \er(13).  But there we are
dealing with nonperturbative quantities.  Hence the off diagonal
terms in the density matrix are explicitly allowed for two
reasons.  First the chiral symmetry is definitely broken in the
nonperturbative part of QCD. Secondly, in the helicity basis,
which is natural to use, the density matrix for the hadron is
itself offdiagonal, so they can be no constraint prohibiting
offdiagonal terms in the quark density matrix.

%=====================================================================
\section {Summary of Proof}%
The proof of a factorization theorem is made by considering a
cross section as the sum over all cut Feynman graphs for the
process in question.  Ideally, one would like to extend the proof
to handle nonperturbative contributions. But that extension has
yet not been made.

The formulation of the theorem is in fact general enough to allow
such an extension, and the physical picture it gives is
completely reasonable.  In particular, the definitions of the
parts of the factorization formulae that are to be used
nonperturbatively, viz, the parton densities, are valid beyond
perturbation theory: The parton densities have a gauge invariant
definition in terms of hadron matrix elements of certain
operators.

Furthermore, in the whole of our discussion, there is no
restriction on which kinds of particle compose the initial state,
except that they should be gauge invariant and physical.  In
perturbative calculations we will typically use on-shell,
physically polarized gluon and quark states, but in principal we
could also use hadron states with a bound-state wave function.
For the general theory, it will make no difference.  The hard
scattering functions are the only quantities for which a purely
perturbative calculation makes sense, and for them the initial
states are always on shell, massless partons.

The steps in the proof \ref[CSS] are:
\item {1.}  Power counting. Apply the method of Libby and
   Sterman\ref[LS] to determine those regions of integration
   momenta of the cut graphs that give leading contributions.
\item {2.}  Cancellation of superleading regions.  There are
   contributions in which all the partons coupling to the hard
   scattering are gluons with an unphysical scalar polarization.
   Unfortunately, these give a power law larger than the final
   result.  Ward identity methods are used to show that these
   contributions exactly cancel among themselves.
\item {3.}  All remaining contributions have a form like that of
   the parton model, with two generalizations.  First, the hard
   scattering is not restricted to be in the Born approximation.
   Secondly, there may be soft gluons connecting the lines
   associated with initial hadrons among themselves and with
   initial and final state lines in the hard scattering, and
   there may be extra collinear gluons with scalar polarization
   connecting to the hard scattering.
\item {4.}  Cancellation of final-state interactions.  All
   interactions that are too late to affect the inclusive cross
   section must cancel.  For example, hadronization of
   final-state jets does not affect the totally inclusive
   structure functions.  Hence the partons initiating these
   jets effectively have virtuality of order $Q^{2}$
\item {5.}  Taylor expansion.  The hard scattering is expanded
   in powers of the small components of its
   external momenta, and in the mass parameters of its internal
   lines.  The subgraphs of collinear lines (`jet subgraphs') are
   expanded in powers of the relatively small components of the
   soft lines that connect them to other jet subgraphs.
\item {6.}  Cancellation of soft gluons.  A Ward identity
   argument is used to factorize the soft gluons, after which a
   unitarity cancellation applies.
\item {7.}  Factorization of collinear scalar gluons.  This
   again goes by a Ward identity argument.
\item {8.}  At this point we have jet factors and hard scattering
   factors.  One now applies combinatoric arguments in the same
   way as in Wilson's expansion to get the factorization theorem
   \ref[Zav,Tk2,JCC].
\item {9.}  The hard scattering can now be identified as a cross
   section for scattering of on-shell partons with subtractions
   to remove the non-ultraviolet contributions.
\item {10.}  Operator definition of parton densities.  The jet
   factors can now be shown to be exactly hadronic expectation
   values of certain operators.  The precise form of these
   operators is completely determined by the Taylor expansion at
   step 5.  The operators are bilocal operators that have simple
   interpretations in light front quantization.  The terms that
   were obtained in step 7 from the collinear scalar gluons
   turn these into gauge invariant operators.

%------
\subsection {Use of Physical Gauge}%
It is possible to use a physical gauge in trying to prove
factorization.  Several of the unpleasant steps involving
gluons with scalar polarization can then be omitted, and the
result appears to be a much simpler proof.

However, there are unphysical singularities that prevent the
unitarity arguments in step 4 from being applied in as strong a
manner as is needed.  Furthermore, the Ward identity arguments to
cancel soft gluons rely on contour deformations that are
obstructed by these same singularities.  Thus it seems best to
work in an ordinary covariant gauge and to accept the
added complications\ref[CSS,CSS1].

%=====================================================================
\section {Power Counting}%
Libby and Sterman\ref[LS] showed that to classify the important
regions of momentum space in a high energy limit it is useful to
measure momenta and masses in units of the large momentum scale
$Q$. Thus one writes a generic momentum and mass in the form
$$
\eqalign{
   k^{\mu } & = Q \tilde k ^{\mu },
\cr
   m & = Q \tilde m.
}
\eqno(21)
$$
By simple dimensional analysis, the large $Q$ limit is equivalent
to the limit of zero mass and on-shell external momenta; for
example a cross section might be written
$$
  \sigma (Q^{2}; m, k) = Q^{-2 }\sigma (1; \tilde m, \tilde k),  \eqno(22)
$$
with $\tilde m \to  0$ and $\tilde k^{2} \to 0$ when $Q\to \infty $
Thus the complications of a high-energy limit can be investigated
by examining the singularities in zero-mass limit.  The method of
Coleman and Norton\ref[CN] shows in a physically appealing
fashion how to determine the configuration of loop momenta that
give these singularities.  Tkachov and collaborators\ref[Tk1,Tk2]
have shown how systematic exploitation of this idea can
considerably streamline proofs of the operator product expansion
and of other results on the asymptotics of Euclidean Green
functions.

In our case, the significant configurations involve internal
lines of three kinds: (a) Lines that carry momenta collinear to
momenta of external particles. (b) Lines that carry soft momenta.
(c) Lines that carry large ultraviolet momenta.  (If only large
momentum lines are important, then the cross section under
discussion is infra-red safe, and the problem may be treated by
classical renormalization-group methods.)

Then the importance of each configuration is determined by
expanding in powers of a suitable small variable $\lambda $ about the
singular point in the massless limit.

Let us now examine these arguments as applied to deeply inelastic
lepton scattering.  We will present them in a sufficiently
general manner, that the extension to other processes will be
simple.

%------
\subsection {Deeply Inelastic Scattering at Twist-2 Level}%
To explain the power counting, it will be convenient to use
light-front coordinates in which
$$
  p^{+}=Q/x\sqrt 2, \ \ p^{-}=M^{2}/Q\sqrt 2, \ \ p_{\perp }=0,
\eqno(23)
$$
and
$$
   -q^{+}=q^{-}=Q/\sqrt 2, \ \ q_{\perp }=0.
\eqno(24)
$$
(Our metric is such that $V^{2} = 2V^{+}V^{-}-V_{\perp }^{2}$ for any vector
$V^{\mu }$.)

Let us also assume that we sum over cuts of graphs for the
structure tensor before applying the Libby-Sterman argument to
deeply inelastic scattering.  The sum over cuts means that no
final-state interactions need enter our argument.

At the leading power of $Q$, contributions to the structure
tensor come from regions symbolized in \fig{2}.  In the upper
part of this diagram, which we will call the hard subgraph $H$,
all the internal lines have large momenta, that is the scaled
momenta have virtuality of order unity.  In the lower part, which
we will call the collinear or jet subgraph, $J$, all the lines
have momenta collinear with $p^{\mu }$.  That is, the corresponding
scaled momenta are close to a light-like vector with only a
nonzero $+$ component.  All but two of the lines joining the
subgraphs are gluons with scalar polarization.

It is a relatively simple generalization\ref[LS] of the arguments
below that shows that graphs with extra quarks and/or
transversely polarized gluons joining the two subgraphs are
suppressed by a power $Q$ for each extra line.

   \deffig{2}{2}{Regions for twist-2 contributions to
      for deep inelastic scattering.}
%   \defpsfigc{2}{2}{Regions for twist-2 contributions to
%      for deep inelastic scattering.}{polf-2.eps}{1000}{1.5}

%---
\subsection {Simple Quark Connection}%
First we consider the case that there is just a single quark line
connecting the subgraphs $H$ and $J$ on each side of the
final-state cut.  As we now show, it is fairly easy to massage
the contribution of \fig{2} into a parton-model-like form
that implies the scaling properties of the structure functions
that were stated earlier.

We must first find the leading part of the trace over Dirac
matrices.  For this purpose, we decompose the top part --- the
hard subgraph $H$ --- and the bottom part --- the jet subgraph
$J$ --- according to the Dirac structure on the fermion line
connecting the two subgraphs:
$$
\eqalign{
  J &= J_{S}1 + J^{V}_{\kappa }\gamma ^{\kappa } + J^{T}_{\kappa \lambda
}\sigma ^{\kappa \lambda } + J^{PV}_{\kappa }\gamma ^{\kappa }\gamma _{5} +
J^{PS}\gamma _{5},
\cr
  H &= H_{S}1 + H^{V}_{\kappa }\gamma ^{\kappa } + H^{T}_{\kappa \lambda
}\sigma ^{\kappa \lambda } + H^{PV}_{\kappa }\gamma ^{\kappa }\gamma _{5} +
H^{PS}\gamma _{5}.
}
\eqno(25)
$$
In the last line we have suppressed the indices $\mu \nu $ of the structure
tensor.  When we perform the trace over the Dirac matrices and the
integral over the explicit loop momentum $k^{\mu }$, we find that:
$$
\eqalign{
  W &= \int \frac {d^{4}k}{(2\pi )^{4}} \, \tr (JH)
\cr
    &= \int \frac {d^{4}k}{(2\pi )^{4}} \,\, 4 \left( J_{S}H_{S} + J^{V}\cdot
H^{V} - 2 J^{T}\cdot H^{T} - J^{PV}\cdot H^{PV} +
       J^{PS}H^{PS} \right).
}
\eqno(26)
$$

We suppose that we are looking only at the region of momentum
appropriate to \fig{2}, so that in particular $k^{+}=O(Q)$, and
$|k^{-}|, |k_{\perp }| \ll Q$. The jet part $J$ depends on the momenta $p^{\mu
}$
and $k^{\mu }$, and on the hadron's spin state defined by $\lambda $ and
$s_{\perp }$.
In the rest frame of the hard scattering, all the vectors
involved in $H$ have components of order $Q$, so that we may
regard all components of the decomposition of $H$ as being of
order $Q^{D}$, where $D$ is the mass dimension of $H$.

As we increase $Q$ while keeping the longitudinal momentum fraction,
virtuality and transverse components of $k^{\mu }$ fixed, we may consider $J$
as being obtained by a boost from the rest frame of $p^{\mu }$.
Then the terms in the decomposition of $J$ scale with $Q$ as
follows:
$$
\eqalign{
                 J_{V}^{+}, J_{A}^{+}, J_{T}^{+i} & \propto  Q^{1}, \cr
  J_{V}^{i}, J_{A}^{i}, J_{T}^{+-}, J_{T}^{ij}, J_{S}, J_{PS} & \propto  Q^{0},
\cr
                J_{V}^{-}, J_{A}^{-}, J_{T}^{-i}  & \propto  Q^{-1},
}
\eqno(27)
$$
the proof following from the effect of boost transformations. The
indices $i$ and $j$ refer to purely transverse components.

It follows that the leading terms in the trace in \eq(26) are given by
$$
\eqalign{
  W = & \int \frac {\d^{4}k}{(2\pi )^{4}} \,\, 4 \left( J^{V+}H^{V-}- 2
J^{T+i}H^{T-i} - J^{PV+}H^{PV-} \right)
\cr
      & \left( 1 + O(\hbox{mass}/Q\times \hbox{logarithms}) \right).
}
\eqno(28)
$$

%------
\subsection {Relation to Quark Distribution}%
In \fig{2} we aim to associate the subgraph $H$ with a
contribution to the hard scattering coefficient and the subgraph
$J$ with contribution to a parton distribution.  Since all
components of momentum inside $H$ are of order $Q$, we may,
within $H$, neglect the transverse momentum and virtuality of
$k^{\mu }$ and write
$$
  W = \int  \d\xi  \tr H(q,\xi ) \left[ p^{+} \int \frac {dk^{-} \,
d^{2}k_{\perp }}{(2\pi )^{4}}\, J(k,p,s) \right]
     + {\rm nonleading\ power},
\eqno(29)
$$
where $\xi \equiv k^{+}/p^{+}$, while $H$ has been approximated by something
with
an incoming onshell quark that has zero transverse momentum.  We
also set the quark masses in $H$ to zero.

It remains to discuss the polarization structure.

In the frame we have chosen, it is manifest that the leading
power for $\tr JH$ comes from the terms in \eq(28).  To relate
this to a standard spin projection for Dirac particles, recall
the conventional projection for a spinor wave function:
$$
  (p\llap /+m)\, (1+\gamma _{5}s\llap /).
\eqno(30)
$$
This is singular in the zero mass limit. After application of the
decomposition \er(2) in terms of helicity and transverse spin, an
expression is obtained that has a well-behaved zero-mass limit:
$$
    p\llap / \, \left( 1 + \gamma _{5}s\llap /_{\perp } - \lambda \gamma _{5}
\right).
\eqno(31)
$$

After some reorganization of the $\gamma $ matrices implicit in
\eq(28), we get a contribution of the form
$$
  W_{\mu \nu } = \int _{x}^{1} \frac {d\xi }{\xi } \frac {1}{2}\tr \overline
H_{\mu \nu } \hat k\llap / [1+\gamma _{5}(\lambda _{q}+s\llap /_{q\perp })]
f(\xi )
        + \hbox {twist higher than 2}.
\eqno(32)
$$
where $\overline H$ means the massless on-shell limit of $H$, and
$\hat k^{\mu }\equiv (\xi p^{+},0,0_{\perp })$.  We have restored the $\mu \nu
$ indices that
correspond to the external photon.

The quantity $f(\xi )$ in this equation represents the contribution
of the lower part of \fig{2} to the quark density.  We need to sum
over all possible graphs and all possible regions, and to apply
the same combinatoric arguments as for the operator product
expansion.  Then we should expect to get the following definition
of the quark density:
$$
  f(\xi ) = \int \frac {dk^{-} \, d^{2}k_{\perp }}{(2\pi )^{4}}\, \tr J(k,p,s)
\frac {\gamma ^{+}}{2},
\eqno(33)
$$
while the quark helicity, $\lambda _{q}$, and quark transversity, $s_{q\perp
}$,
are defined by
$$
  \lambda _{q}f(\xi ) = \int \frac {dk^{-} \, d^{2}k_{\perp }}{(2\pi )^{4}}\,
\tr J(k,p,s) \frac {\gamma _{5}\gamma ^{+}}{2},
\eqno(34)
$$
and
$$
  s_{q\perp }f(\xi ) = \int \frac {dk^{-} \, d^{2}k_{\perp }}{(2\pi )^{4}}\,
\tr J(k,p,s) \frac {\gamma ^{+}}{2}\gamma _{5}\gamma _{\perp }.
\eqno(35)
$$
The reasoning given above for these to be appropriate definitions
can be found in \ref[RS], where references to earlier work on
unpolarized parton densities in light front quantization can be
found.

Equation \er(32) is clearly of the form of the desired factorization
theorem. Moreover, the trace with $\overline H$ in \eq(26) is such that the
result has the normalization of the structure tensor for deep inelastic
scattering off a quark target with momentum $\xi p$, helicity $\lambda _{q}$
and
transverse spin $s_{q\perp }$.  Furthermore $f_{i}(\xi )$ must be interpreted
as the
number density of partons.  (The factor $1/\xi $ in \eq(32) is
needed to interpret $f(\xi )$ as a number density because of the
relativistic normalization of the states.)  There will be
technicalities to generalize these results to real QCD, but the
power counting arguments will remain unchanged.

%---
\subsection {Transverse Polarization}%
A twist-2 contribution to the structure functions
$(M/\sqrt {p\cdot q})g_{2}$ and $g_{1}$ is one that is of order $Q^{0}$ (modulo
the
usual logarithms), since they are dimensionless: all the factors
in the tensors multiplying them in \eq(7) are dimensionless
ratios of momenta that are of order $Q$.  The power counting
argument just presented shows that our basic expectation is that
$(M/\sqrt {p\cdot q})g_{2}$ scales like $Q^{0}$.

(In doing the power counting, we consider in the first instance,
the structure tensor $W_{\mu \nu }$, which is dimensionless.  Then we
derive results for the structure functions by considering
the possible tensors in \eq(5), but treating the tensors in
combinations that scale as $Q^{0}$ in the Bjorken limit.  The
coefficients of these tensors then also scale as $Q^{0}$ (times
logarithms).  The unpolarized structure functions $F_{1}$ and $F_{2}$
are such coefficients.  Since the spin vector for a
longitudinally polarized proton can be taken as $\lambda p^{\mu }/M$, the
coefficient of $g_{1}$ is $i\lambda \epsilon _{\mu \nu \rho \sigma }q^{\rho
}s^{\sigma }/p\cdot q$, so that $g_{1}$ scales as
$Q^{0}$.  But the part of the spin vector that goes with $g_{2}$ is the
transverse part $s_{\perp }^{\mu }$, which is invariant when one goes to the
Bjorken limit; thus it is the combination $(M/\sqrt {p\cdot q})g_{2}$ that
must be considered in our scaling argument.)

Now, when one actually performs the calculation of Feynman
graphs to the leading power of the hard subgraphs $H$, one gets
zero for the part corresponding to $g_{2}$.  The most basic way of
seeing this is to observe that the leading power of $Q$ is given
by inserting massless propagators everywhere in the hard part.
Then $H$ must contain an odd number of Dirac matrices, and this
gives zero in the trace with $k\llap /\gamma _{5}s\llap /_{p\perp }$ in
\eq(32). (Similar
reasoning shows that the other two terms in \eq(32) are generally
nonzero.) This argument works to all orders of perturbation
theory, and demonstrates that $(M/\sqrt {p\cdot q})g_{2}$ is suppressed by at
least one power of $Q$.  Since the first nonleading power term is
presumably nonzero, it is in fact $g_{2}$, with the conventional
definition, that scales.

A fancier way of saying the same thing is to observe that in the
zero mass limit, both QCD and the electromagnetic vertices are
chirally invariant.  Chiral invariance prohibits helicity
flip for the quarks, and $g_{2}$ corresponds to an offdiagonal term
in the density matrix, so that it necessarily involves helicity
flip.

If one calculates Feynman graphs for $H$, using an on-shell
projection, but leaving the quark masses nonzero, then the result
is of course proportional to the quark mass.  However, to treat
this as the dominant contribution to $g_{2}$ is an entirely
incorrect application both of parton model ideas and of QCD.  In
the first place, as the above argument makes clear, there are
other terms in the projection over Dirac matrices besides the
ones that give the twist-2 terms.  These terms cannot be
interpreted as the product of the scattering of on-shell quarks
times a quark number density.  Rather they correspond to coupling
to the matrix elements of the twist-3 operators that were listed
by Jaffe and Ji\ref[JJ]. Furthermore, there are regions other
than those of \fig{2} that contribute at the twist 3 level; the
corresponding operators involve, for example, the correlation of
a gluon with quarks\ref[QS,BB].

When one attempts to force a connection in the standard fashion
between a transverse spin dependence of the quark densities and
the transverse structure functions, contradictions
arise\ref[IKL,F].  These have given an undeserved idea in the
folklore that transverse spin cannot be treated within the parton
model and its QCD realization.

It should also be clear that if one represents the size of the
twist 3 contributions as being of order $M/Q$ relative to a
typical twist 2 term, then $M$ should be some kind of hadronic
mass scale, hundreds of MeV, at the least.  The effect of putting
a current quark mass into the calculation of the hard subgraph is
a rather small effect, and hardly can be expected to be the
dominant nonleading contribution.

%---
\subsection {Most General Case}%
To the extent that the gauge properties of QCD are irrelevant,
the argument given above can be readily turned into a full proof.
(One has a minor generalization that it is necessary also to
consider the possibility of gluon lines joining the jet and hard
subgraphs.)

Moreover, the argument can be further generalized, for example to
processes like Drell-Yan with two hadrons in the initial state.
If both partons are both transversely polarized, then the
helicity conservation argument no longer prohibits a transverse
spin dependence of the cross section for such a process, rather
the contrary.

Our argument shows clearly that there is no problem in defining
the concept of a transversely polarized quark.  Such an argument
was (to my knowledge) first constructed, in the context of the
Drell-Yan process, by Ralston and Soper\ref[RS]. In that process,
transversely polarized quarks do indeed give contributions to the
cross section, at the level of twist-2 terms.  Ralston and Soper
were working at a time before the full proof of the factorization
theorem had become worked out.

However, as far as proofs of factorization in a gauge theory are
concerned, there are three essential complications.  First, if
there is a quark connection, then it is possible to have extra
gluons connecting $H$ and $J$. Second, it is possible to have
Faddeev-Popov ghost connections, for which there is no physical
parton distribution.  Third, if there are only gluons connecting
the two subgraphs, then the leading power is $Q^{2}$ times the
canonical power.  We will treat these complications in the next
section.  The important point is to see that these issues are
the same as in unpolarized scattering, so that we need only quote
previous results.

%=====================================================================
\section {Super-leading Terms}%
Consider now the case of a gluon connecting the two subgraphs in
\fig{2}.  We may represent this by
$$
  J^{\alpha ...} (-g_{\alpha \beta }) H^{\beta ...}, \eqno(36)
$$
where the dots ($...$) represent the indices for the other lines
and for the virtual photons attached to $H$.  By exactly the same
argument as in the previous subsection, the leading power in
\er(36) comes from
the term
$$
  -\frac {J^{+}}{p^{+}} p^{+}H^{-}.  \eqno(37)
$$
We have multiplied and divided by $p^{+}$ to exhibit a jet factor
that is boost invariant.  It is easy to check if all the lines
joining the two subgraphs are gluons, then we get a contribution
to the structure functions that is a factor of $Q^{2}$ larger than
the twist-2 contribution that we got in the quark case.  We call
this contribution `super-leading'.

Moreover, if we start with a contribution with just a pair of
quark lines, and if we add an extra gluon line joining the top
and bottom, then the term \er(37) in the gluon polarization gives
a contribution that has the same power law as before the gluon
was added.  This is specific to the case of a vector field.  If
we add an extra fermion or an extra scalar line (in the case that
we have a model with elementary spin zero fields), then the
numerator factors are insufficient to compensate the extra large
denominator in $H$, and we lose a power of $Q$.  The factor
\er(37) results in a factor $Q$ greater than `normal'.

To handle these contributions, we make the following
decomposition of the numerator $-g_{\alpha \beta }$ of the gluon propagator in
\er(36):
$$
\eqalign{
  -g_{\alpha \beta } &= \frac {n_{\alpha }k_{\beta }}{n\cdot k} + \left(
-g_{\alpha \beta } + \frac {n_{\alpha }k_{\beta }}{n\cdot k} \right)
\cr
       &= \frac {n_{\alpha }k_{\beta }}{n\cdot k} + h_{\alpha \beta }.
}
\eqno(38)
$$
Here, $k^{\alpha }$ is the momentum of the gluon, supposed collinear to
$p^{\mu }$, and $n^{\alpha }$ is a vector with just a $-$ component: $n^{\alpha
}=\delta _{-}^{\alpha }$.
(This vector gives a covariant definition of the fractional
momentum carried by $k$: $\xi =n\cdot k/n\cdot p$.)  The leading term \er(37)
is entirely contained in the first term in \eq(38), while the
contribution given by $h_{\alpha \beta }$ is exactly one power of $Q$ smaller.

When we expand each of the gluons between $H$ and $J$ by using
\eq(38), we will say that the gluons with the $n_{\alpha }k_{\beta }/n\cdot k$
term
have scalar polarization, while those with the $h_{\alpha \beta }$ term have
transverse polarization.

The largest superleading term is obtained when all the
connections between $H$ and $J$ are gluons with the $n_{\alpha }k_{\beta
}/n\cdot k$
term.  Since every gluon attached to $H$ is given a factor $k^{\beta }$,
there is a cancellation\ref[sl] by a Ward identity.  This
cancellation is exact in an abelian theory.  But since we must
exclude graphs that are one-particle-reducible in each group of
collinear external lines, we obtain extra terms in a nonabelian
theory.  For example, there are commutators between the different
scalar gluons.  The same argument must be applied recursively to
the commutators.  The details of such an argument have never been
worked out in detail, even in the unpolarized case, to the best
of my knowledge.  In any event, the argument has nothing to do
with the polarization of any of the external particles involved.

When there is only one transverse gluon, with the other
connecting lines being scalar gluons, the same argument
applies.  The transverse gluon gives some nonzero terms in the
Ward identity, involving the transverse gluon.  But there is no
connection on other side of the final state, which means the
process cannot happen.  In a nonabelian theory, there should also
be terms in the Ward identity that bring in ghosts and that
cancel the contributions when Faddeev-Popov ghosts connect the
hard and jet subgraphs.  Again, the polarization state of the
initial hadron is entirely irrelevant.

We are therefore left with contributions that involve either two
transverse gluon lines or two quark lines, one on each side of
the final-state cut, together with arbitrarily many scalar
gluons.  The identical argument given above is used to extract
the leading term for the quark lines, then a Ward identity is
used to move the scalar gluons to the quarks, and thereby
build up a gauge-invariant quark operator.  The case of two
transverse gluons is handled similarly:  Only values of the index
$\beta $ in $h_{\alpha \beta }$ that are in the transverse plane give a leading
contribution, and the combination of one of these factors on each
side of the cut is exactly what corresponds to the spin density
matrix for a gluon.

%=====================================================================
\section {Extraction of Soft Gluons}%
Deeply inelastic scattering is special: there are no soft gluons
to worry about after the sum over final-state cuts.

%=====================================================================
\section {Factorization}%
We now need combinatoric arguments to go from the decomposition
given above, region-by-region, to the factorization \eq(7) with
an operator definition of the parton densities.  These arguments
are of exactly the same form as for Wilson's original operator
product expansion in the Euclidean case\ref[Zav,Tk2,JCC].  Once
one has established the leading regions to be those symbolized by
\fig{2}, the fact of having a different kinematic definition of
the regions is irrelevant.

Thus we may regard \fig{2}, without the extra scalar
gluons, as being the factorization.  The spin structure is
entirely contained in the Dirac structure we elucidated above.
That enables us to read off the correct definitions of the parton
densities.

%=====================================================================
\section {Definition of Parton Distribution Functions}%
Elementary manipulations convert the formula \er(33) for the quark
density into the expectation value of a certain nonlocal
operator\ref[CS]:
$$
  f_{i}(\xi ) = \int  \frac {dy^{-}}{2\pi }\e^{-i\xi p^{+}y^{-}} \langle p| \,
\bar \psi _{i}(0,y^{-},0_{\perp }) \, \frac {\gamma ^{+}}{2} \,
            P \e^{-ig\int _{0}^{y^{-}}dy'^{-}\, A^{+}_{\alpha }(0,y'^{-},0) \,
t_{\alpha }}
            \, \psi _{i}(0) \, |p\rangle .
\eqno(39)
$$
The path ordered exponential is needed to make the operator gauge
invariant, and the manipulations with the Ward identities prove
that it is needed.  When we work in the light-cone gauge $A^{+}=0$,
this exponential vanishes.

The quark helicity and transverse spin are functions of $\xi $.  They are
defined by replacing the $\gamma ^{+}/2$ in \eq(39) by $\gamma _{5}\gamma
^{+}/2$ and $\gamma ^{+}\gamma _{5}\gamma _{\perp }/2$,
respectively.  (See eqs.\ \er(34) and \er(35).)  By conservation
of angular momentum and parity, the quark helicity and transverse
spin are proportional\ref[AM] to the corresponding quantities for
the target, so that we can write
$$
\eqalign{
        \lambda _{i}f_{i}(\xi ) & = \lambda  f_{Li}(\xi ),
\cr
       s_{i\perp }^{\mu }f_{i}(\xi ) & = s_{\perp }^{\mu } f_{Ti}(\xi ),
}
\eqno(40)
$$
where the spin variables with and without the subscript $i$ are
for the parton $i$ and the hadron target, respectively.  (Angular
momentum conservation here refers to the invariance of both the
theory and of definitions like \eq(39) under rotations about the
$z$-axis.

The operator definitions that translate eqs.\ \er(34) and \er(35)
are:
$$
  \lambda  f_{Li}(\xi ) = \int  \frac {dy^{-}}{2\pi }\e^{-i\xi p^{+}y^{-}}
\langle p| \, \bar \psi _{i}(0,y^{-},0_{\perp }) \, \frac {\gamma ^{+}\gamma
_{5}}{2} \,
            P \e^{-ig\int _{0}^{y^{-}}dy'^{-}\, A^{+}_{\alpha }(0,y'^{-},0)\,
t_{\alpha }}
            \, \psi _{i}(0) \, |p\rangle .
\eqno(41)
$$
and
$$
  s_{\perp }^{\mu } f_{Ti}(\xi ) = \int  \frac {dy^{-}}{2\pi }\e^{-i\xi
p^{+}y^{-}} \langle p| \, \bar \psi _{i}(0,y^{-},0_{\perp }) \frac {\gamma
^{+}\gamma _{\perp }^{\mu }\gamma _{5}}{2}
            P \e^{-ig\int _{0}^{y^{-}}dy'^{-}\, A^{+}_{\alpha
}(0,y'^{-},0)t_{\alpha }}
            \psi _{i}(0) \, |p\rangle .
\eqno(42)
$$
The asymmetries $\Delta _{L}=f_{Li}(\xi )/f_{i}(\xi )$ and $\Delta
_{T}=f_{Ti}(\xi )/f_{i}(\xi )$
are, in general, functions of the fractional momentum variable
$\xi $ (and of the scale $\mu $ at which the densities are defined).

Feynman rules are readily written down, as in \fig{3}.  We have diagrams
in which there is an incoming particle for the state $|p\rangle $ on the left,
and an outgoing particle for the state $\langle p|$ on the right.  There is a
cut for the final state.  The operator is represented by the double line
crossing the final state.  Any number of gluons may attach to the double
line.  Integrals over all loop momenta are performed, and in addition
there is an integral over the $k^{-}$ and $k_{\perp }$ coming out of the
vertex;
$k^{+}$ is set equal to $\xi p^{+}$.  Finally the product of Dirac matrices for
the explicit quark line is traced with $\gamma ^{+}/2$ for the unpolarized
density, with $\gamma _{5}\gamma ^{+}/2$ for the helicity part, and with
$\gamma ^{+}\gamma _{5}\gamma _{\perp }/2$ for
the transverse polarization part.  These rules are equivalent to those
written down by Artru and Mekhfi\ref[AM] in a helicity basis.

   \deffig{3}{5}{Feynman rules for quark densities.  }
%   \deftexfigc{3}{polf-3}{Feynman rules for quark densities.  }

The definitions given above have ultra-violet divergences when
$k_{\perp }\to \infty $. These are renormalized in the same way as the
ultra-violet divergences in the ordinary twist-2 local operators.
Modulo the anomalies associated with the $\gamma _{5}$, which have nothing
specific to do with the difficulties in defining transverse
polarization, these renormalizations are straightforward.  The
renormalization procedure introduces explicit dependence on a
renormalization scale $\mu $. The renormalization group equations
for the parton densities are the ordinary evolution equations,
the (Gribov-Lipatov)-Altarelli-Parisi equations.

When integer moments are taken of the above quark densities, and
combined with the appropriate sign (plus or minus) times the
antiquark densities, matrix elements of twist two local operators
are obtained.  For the unpolarized distributions and and for the
helicity distributions, these operators are familiar from the
treatment of deeply inelastic scattering by the operator product
expansion.

%========================================
\section {Covariance of Parton Distributions}
The definitions we have given of the quark distributions depend
on the choice of a frame, and might therefore appear not to be
Lorentz invariant.  So we must now show that this is not so.  The
choice of coordinates can be specified by a vector
$$
  n^{\mu } \equiv  \delta ^{\mu }_{-},                   \eqno(43)
$$
which is lightlike and future pointing.  Then the momentum-space
integrals in \eq(29) and its relatives have a covariant
formulation:
$$
  dk^{-}d^{2}k_{\perp } = d^{4}k \, \delta (k\cdot n - \xi  p\cdot n),
\eqno(44)
$$
while the coordinate space integrals in \eq(39) etc are
$$
  dy^{-} \, \e^{-i\xi p^{+}y^{-}} = d\lambda  \, \e^{-i\xi \lambda p\cdot n}.
         \eqno(45)
$$
The coordinate of the antiquark field in \eq(39) is $\lambda n^{\mu }$, and
the matrix $\gamma ^{+}$ is $\gamma \cdot n$, so that all the definitions are
invariant under boosts along the $z$-axis, that is, they are
invariant under scaling of $n^{\mu }$ to $Cn^{\mu }$.  Finally, the
path-ordered exponential in \eq(39) has the boost invariant form
$$
   P \e^{-ig\int _{0}^{\lambda n} d\lambda ' \, n\cdot A_{\alpha }(\lambda
'n)t_{\alpha }}
\eqno(46)
$$

All the definitions of the parton densities now covariant, and
they all explicitly depend on the fractional momentum variable
$\xi $.  They also depend on the momentum and spin vector of the
hadron $p$, $s$, and on $n$.  The operator expectation values in
the definitions depend linearly on the hadron's density matrix
and so their dependence on $s$ is a constant plus a linear term.

Let us define the hadron's helicity by $\lambda =M s\cdot n / p\cdot n$ and its
transverse polarization by $s_{\perp }^{\mu } = s^{\mu } - \lambda (p^{\mu }/M
- n^{\mu }M/p\cdot n)
= s^{\mu } - p^{\mu } s\cdot n/p\cdot n - n^{\mu }s\cdot nM/p\cdot n^{2}$. The
transverse polarization
satisfies $n\cdot s_{\perp } = 0$.

The distributions for unpolarized quarks and for the helicity
dependence, eqs.\ \er(39) and \er(41) can now be seen to be
Lorentz invariant. Since $n\cdot n=0$ and the distributions are
independent of the scale of $n^{\mu }$, the only kinematic variable on
which they can depend (aside from $\xi =k\cdot n/p\cdot n$) is the helicity
$\lambda $, and that at most linearly. Parity invariance of QCD then
shows that the unpolarized distribution is independent of the
hadron polarization, while the helicity asymmetry is linear, just
as we asserted on the left of the defining equations.  (Note that
parity invariance is essential to this result: if QCD did not
conserve parity, then, for example, the number of left handed
quarks in a unpolarized proton need not equal the number of right
handed quarks.)

As defined in \eq(42), the transverse distribution is given as a
Lorentz vector that represents the polarization vector $s_{q}^{\mu }$ of
the quark.  Rotation invariance forces it to be proportional to
the hadron's spin vector, and so the right hand side of \eq(42)
must be a linear combination of $s^{\mu }$, $n^{\mu }s\cdot n/p\cdot n^{2}$ and
$p^{\mu }\lambda $,
with coefficients that are independent of $n$ and of the spin of
the hadron.  We have written the definition in terms of the
transverse components of a gamma matrix $\gamma _{\perp }^{\mu }$, but we could
have
used the complete $\gamma ^{\mu }$ to preserve manifest Lorentz covariance.
The definition satisfies $n\cdot s_{q}=0$, since $(\gamma \cdot n)^{2}=0$. Thus
we get
a linear combination of $s_{\perp }^{\mu }$ and $n^{\mu }s\cdot n/p\cdot
n^{2}$. The coefficient
of $s_{\perp }^{\mu }$ we call the transversity part of the quark density,
$f_{T}(\xi )$.

The term proportional to $n^{\mu }s\cdot n/p\cdot n^{2}$ corresponds to one
of the twist three distribution listed by Jaffe and Ji\ref[JJ],
and arises only if one replaces the $\gamma _{\perp }^{\mu }$ in \eq(42) or
\eq(35)
by $\gamma ^{-}$.  Most importantly, this term gives no contribution in a
twist-2 hard scattering calculation, because one immediately puts
the quark spin into a trace calculation involving the
quantity
$$
    k\llap / (1 - \lambda _{q} \gamma _{5} + \gamma _{5}s\llap /_{q}),
\eqno(47)
$$
where the spin vector is supposed to satisfy $s_{q}\cdot k=0$, with $k^{\mu }$
now being a light-like vector in the + direction.

We have now seen that the definitions we have made of the quark
densities, both the unpolarized one and the longitudinal and the
transversity spin distributions, are explicitly Lorentz
invariant: they are scalar quantities independent of the choice
of the vector $n^{\mu }$.  Essentially the same considerations apply to
the gluon distributions to be defined in the next section.  The
only non-invariance comes into the definition of the hadron
helicity $\lambda $ and the hadron transverse spin vector $s_{\perp }^{\mu }$.

Let us now see why this last noninvariance creates no problem.
The physics is that we define the parton densities to be used in
a conventional twist-2 hard scattering calculation.  The presence
of other particles in the process gives us the definition of
$n^{\mu }$.  Consider for example a collision of two particles of
momenta $p_{1}$ and $p_{2}$---\fig{4}.  Suppose particle 1, which is
moving to the right, has left-handed helicity.  By boosting so
that the reference frame moves faster than the particle, we
reverse its velocity and its helicity, but clearly we have also
changed the view of the collision: particle 2 now overtakes
particle 1.

   \deffig{4}{1.5}{Collision of two
   particles.}
%   \defpsfigc{4}{1.5}{Collision of two
%   particles.}{polf-4.eps}{1000}{1.2}

{}From the point-of-view of the rest frame of particle 1, it is
being probed by a an almost light-like particle moving in a
certain direction.  We can create the exactly light-like vector
$n^{\mu }$ as a linear combination of $p_{2}^{\mu }$ with a {\it small}
admixture of $p_{1}^{\mu }$:
$$
   n^{\mu } \propto  p_{2}^{\mu } - cp_{1}^{\mu },
$$
with $c\approx m_{2}^{2}/s$.  The only ambiguity is that it might be
convenient to choose one of the other momenta in the hard
scattering instead of $p_{2}$ in this formula.  For example, in
deeply inelastic scattering, one often chooses to define
transverse coordinates with respect to $p$ and $q$, which are the
momentum vectors relevant for the hadronic part of the process,
whereas the simpler definition for an experiment is to define
transverse with respect to $p$ and the momentum of the incoming
lepton $l$.  (This is indeed what one means by transverse
polarization in an experiment.)  These two possibilities give
different definitions of the vector $n$ in the parton densities.
In the rest frame of the incoming hadron, they differ by a
rotation through an angle of order $M/\sqrt s$, so that at its largest
the difference corresponds to a twist-3 effect.

%========================================
\section {Operator Definition of Polarized Gluon Distributions}%
Exactly analogous definitions may be made for the density of gluons and
the longitudinal and linear polarization of the gluon.  (A pure state
that is a linear combination of equal amounts of left and right helicity
is called transversely polarized for a spin-\half\ particle, but
linearly polarized for a spin-1 particle).

The gauge invariant definitions are
$$
\eqalign{
  f_{g}(\xi )   & = - \sum _{j=1}^{2}\, \int  \frac {dy^{-}}{2\pi \xi
p^{+}}\e^{-i\xi p^{+}y^{-}} \langle p|G^{+j}(0,y^{-},0_{\perp }) \P G^{+j}(0)
|p\rangle ,
\cr
  f_{{\rm hel}g}(\xi )  & = \sum _{j,j'=1}^{2}\, P_{jj'}^{{\rm hel}}\,
          \int  \frac {dy^{-}}{2\pi \xi p^{+}}\e^{-i\xi p^{+}y^{-}} \langle
p|G^{+j}(0,y^{-},0_{\perp }) \P G^{+j'}(0) |p\rangle ,
\cr
  f_{{\rm lin}g}(\xi ) &  = \sum _{j,j'=1}^{2}\, P_{\perp ,jj'}^{{\rm lin}}\,
      \int  \frac {dy^{-}}{2\pi \xi p^{+}}\e^{-i\xi p^{+}y^{-}} \langle
p|G^{+j}(0,y^{-},0_{\perp }) \P G^{+j'}(0) |p\rangle ,
}
\eqno(48)
$$
where $G_{\mu \nu }$ is the gluon field strength tensor and $\P$ denotes the
path-ordered exponential of the gluon field along the light-cone that
makes the operators gauge-invariant, in exact analogy to \eq(39):
$$
  \P = \exp \left[ \int _{0}^{y^{-}} \d y'^{-} \, A^{+}_{\alpha
}(0,y'^{-},0_{\perp }) T_{\alpha } \right].
\eqno(49)
$$
Here $T_{\alpha }$ are the generating matrices for the adjoint representation
of
color SU(3).  The $j$ index runs over the two transverse dimensions, and
the spin projection operators are defined by
%  ??=====================CHECK conventions
$$
\eqalign{
   P_{11}^{{\rm hel}}   =  P_{11}^{{\rm hel}}  = 0,
\cr
   P_{12}^{{\rm hel}}   =  -P_{21}^{{\rm hel}}  = i,
\cr
   P_{n,jj'}^{{\rm lin}} = n_{j}n_{j'} - \delta _{jj'}/2.
}
\eqno(50)
$$

By angular momentum conservation, the linear polarization of a gluon is
zero in a spin-\half\ hadron\ref[AM].  (The reason is that the linear
polarization is measured by an operator that flips helicity by two
units.  Since no helicity is absorbed by the space-time part of the
definition of the parton densities (the integrals are azimuthally
symmetric), the helicity flip in the operator must correspond to
a helicity flip term in the density matrix for the hadron.

Just as for the quarks, one can take integer moments of the gluon
densities and get matrix elements of local operators.

%=====================================================================
\section {Factorization for Drell-Yan}
There are some complications in the proof of factorization for
Drell-Yan as compared with the proof for deeply inelastic
scattering.  These same complications appear in all other
processes with two hadrons in the initial state. The
complications are explained in \ref[CSS,CSS1], and the proof that
they do not wreck the factorization are entirely independent of
the polarization issue.

The typical leading region corresponds to \fig{1}, which looks
like an obvious generalization of the case of deeply inelastic
scattering, \fig{2}.  There are now two jet subgraphs,
corresponding to the two initial-state hadrons.  There is also
the possibility of extra collinear gluons joining the jet
subgraphs to the hard subgraph, and these are treated in exactly
the same way as in deeply inelastic scattering.  However, and
much more malignantly, there are leading regions in which soft
gluons are exchanged between the two jet subgraphs.  (There are
also soft gluons exchanged with jets going into the final state
from the hard scattering, but these cancel after a sum over
the unobserved part of the final state.)  The soft gluons can be
emitted off internal lines and in the initial state.  Note that a
soft gluon is defined to be one that carries momentum much less
than $Q$, as measured in the center of mass frame.  This is a
broader definition than the one that concerns the very well-known
infrared divergences in QED, and consequently the soft gluons are
not restricted to being emitted from external, onshell colored
particles.

Such interactions were known before QCD: they were called Pomeron
exchange, and physically they generate the observed final states.
The final states corresponding to \fig{1} taken in its naivest
interpretation has two jets of hadrons corresponding to the
remnants of the two incoming hadrons, together with a large gap
in rapidity between them that contains no particles.  This gap is
completely filled in by the Pomeron.

The leading power for the soft interactions is given by a
generalization of the method that generates \er(37) from \er(36).
After that, a proof of cancellation of these soft interactions
involves a tricky combination of Ward identities, analyticity and
unitarity\ref[CSS1].  None of this part of the proof depends on
the polarization.  Note that a key part of the proof rests on
analyticity arguments that were first made in nonperturbative
Pomeron physics\ref[DEL]: these involve analyticity (i.e.,
causality) but not polarization, and are valid to the whole
leading power.

Once one has got the soft interactions canceled, and the
collinear scalar gluons factored out of the hard part, one
is left with the task of determining the leading power part of
the traces over Dirac matrices joining the jet subgraphs and the
hard subgraph.  (There is the same task to perform for gluons.)
This just involves two copies of the argument given above for
deeply inelastic scattering.  Ralston and Soper\ref[RS] gave
this argument before the full apparatus of factorization was
formulated, and we now see their argument must be true in full
QCD.

The only difference from deeply inelastic scattering is that the
suppression of transverse polarization no longer occurs.  If we
have both initial hadrons transversely polarized, then there is a
twist-2 asymmetry in the hard scattering cross section that is
explicitly nonzero at the Born graph level---an asymmetry that is
well known in $e^{+}e^{-}$ physics.

One error does occur in the Ralston-Soper paper.  They attempt to
list all the rather numerous structure functions that are
permitted in the decomposition of the dilepton angular
distribution, and they miss some  (These are not structure
functions in the common misusage of the term.)  The error was
corrected by Donoghue and Gottlieb\ref[DG], and does not at all
effect the general principles.  In any case, only one of the
polarized structure functions is actually nonzero for the Born
graph.

%=====================================================================
\section {Factorization for Other Processes}%
In this paper, I have restricted attention to inclusive processes
with a single large scale.  The methods apply to many processes.
Although not explicitly treated here, the issues in handling the
single particle distribution in a jet are isomorphic to those in
the distribution of partons in a hadron.  So the methods apply
equally to high $p_{\perp }$ single-particle production in hadron-hadron
collisions, to inclusive particle production in $e^{+}e^{-}$
annihilation, or to inclusive particle production in deeply
inelastic lepton scattering, for example.  It would be useful to
make a complete characterization of the processes for which a
factorization theorem of the standard type holds.  All the
considerations of the present paper apply to any of these
processes.

One interesting possibility is that of measuring the correlation
between two particles in a jet.  This should be correlated with
the spin of the parton that initiates the jet, and may provide a
useful handle to probe the transverse spin
distribution\ref[trfrag].

There are many situations in which there is a second scale
associated with the hard scattering.  The simplest is the
Drell-Yan process when $q_{\perp } \ll Q$.  A full factorization theorem
has been stated\ref[qtdy] for this process, and has been
proved\ref[qtepem] for the analogous processes of two-particle
production and the energy-energy correlation in $e^{+}e^{-}$
annihilation in the back-to-back region.  It would be
interesting, and not too hard, to extend these theorems to the
polarized case.  The polarization-specific issues are orthogonal
to and decoupled from the complications associated with the low
$q_{\perp }$ region.

Another case is where the hard scattering has a scale $Q$ much
less than the center-of-mass energy $\sqrt s$.  This is called the
semi-hard region or the small $x$ region.  A leading logarithm
statement of a kind of factorization has been stated by Lipatov
and coworkers\ref[Lip].  The proof leaves much to be desired, and
does not go beyond the leading logarithm level.  Much work
remains here.

Another area is that of exclusive processes.  The state of the
the factorization theorems and their proofs is not nearly so good
as for inclusive processes.  There are considerable
complications\ref[excl] because of regions other than the
simplest short-distance scattering: for example in hadron-hadron
elastic scattering at large angle, there is competition between
the short-distance scattering and the Sudakov-suppressed
Landshoff process. Disentangling the polarization dependence
would be interesting as a piece of theory, but may not be a high
priority because of the minute cross sections at the high values
of $Q$ where perturbative methods unambiguously apply.

Perhaps the most interesting recent developments in the theory of
polarized hard scattering have been the realization\ref[QS,BB] that
a generalization of the factorization theorem appears to be
provable for the first nonleading twist.  Now the first
nonleading twist that is relevant in polarized scattering, with
transversely polarized beams, is twist 3: that is, the first power
corrections are a single power $1/Q$ down from the twist-2 terms.
In many cases of single spin asymmetries, the twist-2 term
vanishes.  An interesting phenomenology should result.  Qiu and
Sterman\ref[QS] have explained the validity of factorization in this
case; Jaffe and Ji\ref[JJ] have listed the operators that are
needed to define the single-body parton distributions.
Interesting physics also lies in two parton correlations that are
an essential parton of the twist-3 results.  There are
experiments on the single transverse spin asymmetries of single
particle production at relatively large $p_{\perp }$ that are greatly in
need of theoretical interpretation.

%=====================================================================
\section {Conclusions}%
The factorization theorems for hard scattering are as true when
the incoming hadrons are polarized as when they are unpolarized.
This is also true for processes in which one measures the
polarization of hadrons in the final state (from fragmentation of
a jet).

The parton densities have unambiguous definitions, which are just
matrix elements of gauge invariant generalizations of the quark
and gluon number operators that are natural in light-front (or
infinite-momentum) quantization.  In the case of polarized beams,
the operators are just those that are directly related to a spin
density matrix for the partons.

This has particular consequences:  The quark transversity
distribution that is relevant for twist-2 processes with
transversely polarized hadrons is perfectly well defined,
contrary to what one might conclude from a superficial reading of
the literature\ref[F,IKL].  The helicity asymmetry of the gluon
density is well defined.  Its first moment is in general a
nonlocal operator, unless one uses the light cone gauge $A^{+}=0$.

%=====================================================================
\mainhead {Acknowledgments}%
This work was supported in part by the U.S. Department of Energy
under grant DE-FG02-90ER-40577, and by the Texas National
Laboratory Research Commission.  I would like to thank many
colleagues for discussions, notably, S. Heppelmann, R.L. Jaffe,
X.-D. Ji, A.H. Mueller, D. Soper, and G. Sterman.

%=====================================================================
\mainhead {Reference}
\defref{AM}{X. Artru and M. Mekhfi, \zphc \rf{45}{669}{90}.  }

\defref{BB}{I.I. Balitsky and V.M. Braun, \np \rf{B311}{541}{89}.}

\defref{BD}{J.D. Bjorken and S.D. Drell, ``Relativistic Quantum Fields''
(McGraw-Hill, New York, 1966).  }

\defref{CN}{S. Coleman and R. Norton, Nuovo Cim.\
   \rf{28}{438}{65}.  }

\boxref \PSUREF \PSUBOX {\ref[PSU]}
\defref{controv}{See for example some of the articles in \PSUREF.}

\defref{CS}{J.C. Collins and D.E. Soper, \np \rf{B194}{445}{82},
   and references therein.
   % parton distribution and decay functions.
   }

\defref{CSS}{J.C. Collins, D.E. Soper and G. Sterman,
   ``Factorization of Hard Processes in QCD''
   in ``Perturbative QCD" (A.H. Mueller, ed.) (World
   Scientific, Singapore, 1989), and references therein.  }

\defref{CSS1}{J.C. Collins, D.E. Soper and G. Sterman, \np
   \rf{B261}{104}{85} and \rf{B308}{833}{88}; G.T. Bodwin, \pr
   \rf{D31}{2616}{85} and \rf{D34}{3932}{86}.  }

\defref{DEL}{C. DeTar, S.D. Ellis, and P.V. Landshoff, \np
   \rf{B87}{176}{75}; J. Cardy and G. Winbow, \pl
   \rf{52B}{95}{74}.}

\defref{DG}{J.T. Donohue and S. Gottlieb, \pr
   \rf{D23}{2577,2581}{83}.}

\defref{excl}{Yu.L. Dokshitzer, V.A. Khoze, A.H. Mueller and
   S.I.Troyan, ``Basics of Perturbative QCD'' (Editions
   Fronti\`eres, Gif-sur-Yvette, 1991), give a review.}

\defref{F}{R. Feynman, ``Photon-Hadron Interactions'', (Benjamin,
   Reading, 1972).}

\defref{hel}{R. Gastmans and T.T. Wu, ``The Ubiquitous Photon'',
   (Oxford University Press, 1990).  This reviews helicity
   methods and gives a very comprehensive list of helicity
   amplitudes that have been calculated. }

\defref{HighSpin}{P. Hoodbhoy, R.L. Jaffe, and A.V. Manohar,
   \np\rf{B312}{571}{89};
   R.L. Jaffe and A.V. Manohar, \np \rf{B321}{343}{89};
   R.L. Jaffe and A.V. Manohar, \pl \rf{223B}{218}{89}.}

\defref{IKL}{B.L. Ioffe, V.A. Khoze and L.N. Lipatov, ``Hard
   Processes, Vol.\ 1'' (North-Holland, Amsterdam, 1984), p.\
   253.  }

\defref{JCC}{J.C. Collins, `Renormalization' (Cambridge University
   Press, Cambridge, 1984).  }

\defref{JJ}{R.L. Jaffe and X.-D. Ji, \prl \rf{67}{552}{91}.
%   \pr \rf{D43}{724}{91}.
}

\defref{Lip}{L.N. Lipatov, in ``Perturbative QCD" (A.H. Mueller,
   ed.) (World Scientific, Singapore, 1989), and references
   therein.  }

\defref{LS}{S. Libby and G. Sterman, \pr \rf{D18}{3252,4737}{78}.}

\defref{PSU}{Proceedings of the Polarized Collider Workshop,
   (J.C. Collins, S. Heppelmann and R. Robinett, eds.) pages 1--9
   (A.I.P., New York, 1991).}

\defref{QS}{J.-W. Qiu and G. Sterman, \np \rf{B353}{105,137}{91}
   and \prl \rf{67}{2264}{91}. }

\defref{RS}{J.P. Ralston and D.E. Soper, \np \rf{B152}{209}{79}.  }

\defref{qtepem}{J.C. Collins and D.E. Soper,, \np
   \rf{B193}{381}{81}.}

\defref{qtdy}{J.C. Collins, D.E. Soper and G. Sterman, \np
   \rf{B250}{199}{85}.}

\defref{sl}{J.M.F. Labastida and G. Sterman
   \np\rf{B254}{425}{85}.}

\defref{trfrag}{J.C. Collins, S. Heppelmann, Bob Jaffe, X.-D. Ji
   and G. Ladinsky, ``Measuring Transversity Densities in Singly
   Polarized Hadron-Hadron Collisions'' preprint PSU/TH/101, in
   preparation.}

\defref{trfr-sgl}{J.C. Collins, ``Fragmentation of Transversely
   Polarized Quarks Probed in Transverse Momentum
   Distributions'', Penn State preprint PSU/TH/102, in
   preparation.}

\defref{Tk1}{F.V. Tkachov, ``Euclidean Asymptotic Expansions of
   Green Functions of Quantum Fields (I) Expansions of Products
   of Singular Functions'', Fermilab preprint
   FERMILAB-PUB-91/347-T.  }

\defref{Tk2}{G.B. Pivovarov and F.V. Tkachov, ``Euclidean
   Asymptotic Expansions of Green Functions of Quantum Fields (II)
   Combinatorics of the $As$-Operation'', Fermilab preprint
   FERMILAB-PUB-91/345-T.  }

\defref{Zav}{O.I. Zavialov, ``Renormalized Quantum Field
   Theory'' (Kluwer, Boston, 1989).}

\outrefs

\bye